\documentclass[12pt]{article}

\usepackage{natbib}
\usepackage[nottoc,notlof,notlot]{tocbibind} 

\bibpunct{(}{)}{,}{a}{}{;}
\usepackage{graphicx}
\graphicspath{ {images/} }

\usepackage{amsmath}
\usepackage{bbm}
\usepackage{amsfonts}%
\usepackage{amssymb}%

\usepackage{geometry} 
\usepackage{graphicx} 
\usepackage{framed}
\usepackage{caption}
\usepackage{subcaption}

\usepackage{booktabs} 
\usepackage{array} 
\usepackage{paralist} 
\usepackage{verbatim} 

\usepackage{lipsum}
\usepackage{setspace}

\usepackage{lipsum}
\usepackage[singlelinecheck=false]{caption}
\usepackage{tabularx}

\usepackage[colorlinks]{hyperref}
\hypersetup{allcolors=blue}
\title{\setstretch{1} Identification and Estimation of a Partially Linear Regression Model using Network Data: Inference and an Application to Network Peer Effects}

\author{Eric Auerbach} 

\begin{document}
\maketitle

This paper provides additional results relevant to the setting, model, and estimators of \cite{auerbach2019identification}. Section 1 contains results about the large sample properties of the estimators from Section 2 of \cite{auerbach2019identification}.  Section 2 considers some extensions to the model. Section 3 provides an application to estimating network peer effects. Section 4 shows the results from some simulations.

\section{Large sample results}
This section contains results about the large sample properties of the estimators from Section 2 of \cite{auerbach2019identification}. Results are stated in Sections 1.1 and 1.2. Proofs can be found in Section 1.3. 

\subsection{Large sample results for $\hat{\beta}$}
\subsubsection{Asymptotic normality}
I provide two asymptotic normality results. The first concerns the case when the distribution of $f_{w_{i}}$ has finite support in the sense that $P(||f_{w_i}-f_{w_j}||_{2} = 0) = P(||p_{w_i}-p_{w_j}||_{2} = 0) > 0$ and there exists an $\epsilon > 0$ such that $P(0 < ||f_{w_i}-f_{w_j}||_{2} < \epsilon)= P(0 < ||p_{w_i}-p_{w_j}||_{2} < \epsilon) = 0$. This assumption is satisfied by the blockmodel of \cite{holland1983stochastic} as discussed in Section 2.2.1 of \cite{auerbach2019identification} \citep[see also][]{bickel2013asymptotic}.

\begin{flushleft} 
\textbf{Proposition C1}: Suppose Assumptions 1-4 and $f_{w_i}$ has finite support. Then as $n \to \infty$
\[V_{3,n}^{-1/2}\left(\hat{\beta} - \beta \right) \to_{d} \mathcal{N}\left(0,I_{k}\right)\] where $V_{3,n} = \Gamma_{0}^{-1}\Omega_{0}\Gamma_{0}^{-1}\times s/n$, $\Gamma_{0} = E\left[(x_{i}-x_{j})'(x_{i}-x_{j})|  \hspace{1mm} ||p_{w_{i}}-p_{w_{j}}||_{2} = 0\right]$,  $I_{k}$ is the $k\times k$ identity matrix, and
\begin{align*}
s &= P( ||p_{w_i}-p_{w_j}||_{2} = 0, ||p_{w_i}-p_{w_k}||_{2} = 0)/P( ||p_{w_i}-p_{w_j}||_{2} = 0)^{2} \\
\Omega_{0} &= 4E\left[(x_{i}-x_{j})'(x_{i}-x_{k})(u_{i}-u_{j})(u_{i}-u_{k}) | \hspace{1mm} ||p_{w_i}-p_{w_j}||_{2} = 0, ||p_{w_i}-p_{w_k}||_{2} = 0 \right] \\
\Gamma_{0} &= E\left[(x_{i}-x_{j})'(x_{i}-x_{j})|\hspace{1mm}  ||p_{w_i}-p_{w_j}||_{2} = 0 \right]
\end{align*}
with $u_{i} = \lambda(w_{i}) + \varepsilon_{i}$.
\end{flushleft}

Proposition C1 is included for three reasons. First, it adds to a literature arguing that some of the adverse effects of unobserved heterogeneity may be mitigated when the support of this variation is finite \cite[see also][]{hahn2010panel,bonhomme2015grouped}. Second, the assumption of discrete heterogeneity is not uncommon in empirical work \citep*[see for instance][]{schmutte2014free}. Third, it provides an easy to interpret condition such that $\hat{\beta}$ converges to $\beta$ at the $\sqrt{n}$-rate. 

The second result concerns the more general case when the support of $f_{w_{i}}$ is not necessarily finite. It requires additional structure on the function $f$ and the bandwidth sequence $h_{n}$ given in Assumptions C1 and C2 respectively. 

\begin{flushleft}
\textbf{Assumption C1:} The function$f$ satisfies the $\alpha$-H\"older-continuity condition that there exists $\alpha, C > 0$ such that $\int \mathbbm{1}\left\{ v \in [0,1]: \sup_{\tau \in [0,1]}\left|f(u,\tau)-f(v,\tau)\right| < \varepsilon\right\} dv \geq \left(\frac{\varepsilon}{C}\right)^{1/\alpha}$  for every $u, \varepsilon \in [0,1]$.
\end{flushleft}

Assumption C1 supposes that for any social characteristics $u \in [0,1]$, the mass of other social characteristics with $\varepsilon$-similar linking probabilities to $u$ can be bounded from below by a fractional polynomial of $\varepsilon$. This condition follows if the function $f$ is $\alpha$-H\"older continuous (for every $u, v \in [0,1]$ we have $\sup_{\tau \in [0,1]}\left|f(u,\tau)-f(v,\tau)\right| \leq C|u-v|^{\alpha}$). I do not assume $f$ is $\alpha$-H\"older continuous because it would rule out the blockmodel of Section 2.2.1 as a special case. 


%

\begin{flushleft}
\textbf{Assumption C2:} The bandwidth sequence $h_{n} = C_{7}\times n^{-\rho}$ for $\rho \in \left(\frac{\alpha}{4 + 8\alpha},\frac{\alpha}{2 + 4\alpha}\right)$ and some $C_{7} > 0$. 
\end{flushleft}

The rate of convergence of the bandwidth sequence in Assumption C2 now depends on the exponent from Assumption C1. When $\alpha = 1$ this bandwidth choice is on the order of magnitude used by \cite{ahn1993semiparametric}. 

The second asymptotic normality proof uses Assumption C1 to strengthen Lemma 1. Its proof  can be found in Appendix Section A.3 of \cite{auerbach2019identification}. 
\begin{flushleft} 
\textbf{Lemma A1}: Suppose Assumption C1. Then for any $i,j \in \{1,...,n\}$ 
\begin{align*}
||p_{w_{i}} - p_{w_{j}}||_{2} \leq ||f_{w_{i}} - f_{w_{j}}||_{2} \leq 2 \hspace{1mm} C^{\frac{1}{2+4\alpha}}\left(||p_{w_{i}} - p_{w_{j}}||_{2}\right)^{\frac{\alpha}{1+2\alpha}}
\end{align*}
where $C$ and $\alpha$ are the constants from Assumption C1.
\end{flushleft}  

\begin{flushleft}
\textbf{Proposition C2}: Suppose Assumptions 1-4 and C1-2. Then as $n \to \infty$
\begin{align*}
V_{4,n}^{-1/2}\left(\hat{\beta} - \beta_{h_{n}} \right) \to_{d} \mathcal{N}\left(0,I_{k}\right)
\end{align*}
where $V_{4,n} = \Gamma_{n}^{-1}\Omega_{n}\Gamma_{n}^{-1}/4n$, $\Gamma_{n} = r_{n}^{-1}E\left[(x_{i}-x_{j})'(x_{i}-x_{j})K\left(\frac{\delta_{ij}^{2}}{h_{n}}\right)\right]$, $r_{n} = E\left[K\left(\frac{\delta_{ij}^{2}}{h_{n}}\right)\right]$, $I_{k}$ is the $k\times k$ identity matrix, and
\begin{align*}
\beta_{h_{n}} &=  \beta + \left(r_{n}\Gamma_{n}\right)^{-1}E\left[(x_{i}-x_{j})'(\lambda(w_{i})-\lambda(w_{j}))K\left(\frac{||p_{i}-p_{j}||_{2}}{h_{n}}\right)\right] \\
\Omega_{n} &= \frac{4}{r_{n}^{2}}E\left[\Delta_{i_{1}j_{1}}\Delta_{i_{1}j_{2}}'K\left(\frac{\delta_{i_{1}j_{1}}^{2}}{h_{n}}\right)K\left(\frac{\delta_{i_{1}j_{2}}^{2}}{h_{n}}\right)\right] \\ 
&+ \frac{1}{r_{n}^{2}h_{n}^{2}}E\left[\Delta_{i_{1}j_{1}}\Delta_{i_{2}j_{2}}'K'\left(\frac{\delta_{i_{1}j_{1}}^{2}}{h_{n}}\right)K'\left(\frac{\delta_{i_{2}j_{2}}^{2}}{h_{n}}\right)\left(F_{i_{1}j_{1}t_{1} s_{11}s_{12}} - \delta_{i_{1}j_{1}}^{2}\right)\left(F_{i_{2}j_{2}t_{1}s_{21}s_{22}} - \delta_{i_{2}j_{2}}^{2}\right)\right] \\
&+ \frac{4}{r_{n}^{2}h_{n}^{2}}E\left[\Delta_{i_{1}j_{1}}\Delta_{i_{2}j_{2}}'K'\left(\frac{\delta_{i_{1}j_{1}}^{2}}{h_{n}}\right)K'\left(\frac{\delta_{i_{2}j_{2}}^{2}}{h_{n}}\right)\left(F_{i_{1}j_{1}t_{1}s_{11}s_{12}} - \delta_{i_{1}j_{1}}^{2}\right)\left(F_{i_{2}j_{2}t_{2}s_{11}s_{22}} - \delta_{i_{2}j_{2}}^{2}\right)\right]
\end{align*}
with $\Delta_{ij} = (x_{i}-x_{j})'(u_{i}-u_{j})$, $u_{i} = \lambda(w_{i}) + \varepsilon_{i}$, $\delta_{ij} = \delta(w_{i},w_{j}) = ||p_{w_{i}} - p_{w_{j}}||_{2}$, and $F_{ijts_{1}s_{2}} = f(w_{t},w_{s_{1}})f(w_{t},w_{s_{2}})\left(f(w_{i},w_{s_{1}})-f(w_{j},w_{s_{1}})\right)\left(f(w_{i},w_{s_{2}}) -f(w_{j},w_{s_{2}})\right)$.
\end{flushleft}

Proposition C2 warrants two remarks. First, the variance of the estimator is not necessarily on the order of the inverse of the sample size because the variance of the kernel, $r_{n}^{-2}E\left[K\left(\frac{||p_{i}-p_{j}||^{2}_{2}}{h_{n}}\right)K\left(\frac{||p_{i}-p_{k}||^{2}_{2}}{h_{n}}\right)\right]$, potentially diverges with $n$. One could  remove this term from the variance (that is, set $K\left(\frac{||p_{i}-p_{j}||_{2}}{h_{n}}\right) = r_{n}$ for every $ij$-pair) by choosing a variable bandwidth in which each agent belongs to the same number of matches, though the strategy likely inflates the bias of the estimator relative to $\hat{\beta}$. I leave the study of such an estimator to future work.

The variance is also inflated relative to the infeasible pairwise difference regression using the unknown codegree distances $\{\delta(w_{i},w_{j})\}_{i \neq j}$, due to the variability of the estimated codegree differences $\{\hat{\delta}_{ij}\}_{i \neq j}$ around their probability limits. That variance is given by $ \Gamma_{n}^{-1}\tilde{\Omega}_{n}\Gamma_{n}^{-1}/n$ where
\begin{align*}
\tilde{\Omega}_{n} = \frac{4}{r_{n}^{2}}E\left[\Delta_{i_{1}j_{1}}\Delta_{i_{1}j_{2}}'K\left(\frac{\delta_{i_{1}j_{1}}^{2}}{h_{n}}\right)K\left(\frac{\delta_{i_{2}j_{2}}^{2}}{h_{n}}\right)\right].
\end{align*}
is the first summand of $\Omega_{n}$. 


The second remark is that the asymptotic distribution of $\hat{\beta}$ is not centered at $\beta$, but at the pseudo-truth $\beta_{h_{n}}$. Though $\beta_{h_{n}}$ converges to $\beta$, the rate of convergence can be slow depending on the size of $\alpha$ and the rate of convergence of $\sup_{u,v \in [0,1]: ||f_{u}-f_{v}||_{2} \leq h_{n}}\left(\lambda(u)-\lambda(v)\right)^{2}$. This bias is standard with matching estimators \citep[see for instance][]{abadie2006large}, although the problem is exacerbated here by the relatively weak relationship between the codegree and network distances given by Lemma A1.  Accurate inferences about $\beta$ using Proposition C2 will generally require a bias correction. 

\subsubsection{Bias correction}
I propose a variation on the jackknife technique proposed by \cite*{honore1997pairwise}, which relies on two additional regularity conditions.
\begin{flushleft}
\textbf{Assumption C3:} The pseudo-truth function $\beta_{h}$ satisfies
$\beta_{h}  = \sum_{l=1}^{L}C_{l}h^{l/\theta} + O\left(h^{(L+1)/\theta}\right)$
for some positive integer $L > \alpha/(2\theta(1 + 2\alpha))$, $k$-dimensional constants $C_{1}, C_{2}, ..., C_{L}$, $\theta > 0$, and $h$ in a fixed open neighborhood to the right of $0$. The matrix  $\Gamma_{n} = r_{n}^{-1}E\left[(x_{i}-x_{j})'(x_{i}-x_{j})K\left(\frac{\delta_{ij}^{2}}{h_{n}}\right)\right]$ converges to the limit $\Gamma_{0}$ as $n\to \infty$.  
\end{flushleft}

Assumption C3 assumes that the pseudo-truth $\beta_{h_{n}}$ can be well approximated by a series of fractional polynomials and that the denominator term $\Gamma_{n}$ has a large sample limit. The method produces a bias-corrected estimator $\bar{\beta}_{L}$. For an arbitrary sequence of distinct positive numbers $\{c_{1},c_{2}, ..., c_{L}\}$ with $c_{1} = 1$, $\bar{\beta}_{L}$ is defined to be 
\begin{align}\label{biascorrected}
\bar{\beta}_{L} = \sum_{l=1}^{L}a_{l}\hat{\beta}_{c_{l}h_{n}}
\end{align}
in which $\hat{\beta}_{c_{l}h_{n}}$ refers to the pairwise difference estimator (4) with the choice of bandwidth $c_{l} \times h_{n}$ and the sequence$\{a_{1},a_{2},...a_{L}\}$ solves
\[ \left( \begin{array}{cccc}
1 & 1 & ... & 1 \\
1 & c_{2}^{2/\theta} & ... & c_{L}^{2/\theta} \\
\vdots & \vdots & \ddots & \vdots \\
1 & c_{2}^{L/\theta} & ... & c_{L}^{L/\theta} \end{array} \right)
\times
 \left( \begin{array}{c}
a_{1} \\
a_{2} \\
\vdots \\
a_{L} \end{array}
 \right)
 =
 \left( \begin{array}{c}
1 \\
0 \\
\vdots \\
0 \end{array}
\right).
\]
\newline

\begin{flushleft}
\textbf{Proposition C3}: Suppose Assumptions 1-4 and C1-C3. Then as $n\to \infty$
\begin{align*}
V_{5,n}^{-1/2}\left(\bar{\beta}_{L} - \beta \right) \to_{d} \mathcal{N}\left(0,I_{k}\right)
\end{align*} 
where $V_{5,n} = \sum_{l_{1}=1}^{L}\sum_{l_{2}=1}^{L}a_{l_{1}}a_{l_{2}}\Gamma_{0}^{-1}\Omega_{n,l_{1}l_{2}}\Gamma_{0}^{-1}/n$, $\Gamma_{0} = \lim_{n\to\infty} r_{n}^{-1}E\left[(x_{i}-x_{j})'(x_{i}-x_{j})K\left(\frac{\delta_{ij}^{2}}{h_{n}}\right)\right]$, $r_{nl} = E\left[K\left(\frac{\delta_{ij}^{2}}{c_{l}h_{n}}\right) \right]$, $I_{k}$ is the $k\times k$ identity matrix, and
\begin{align*}
\Omega_{n,l_{1}l_{2}} &= \frac{4}{r_{nl_{1}}r_{nl_{2}}}E\left[\Delta_{i_{1}j_{1}}\Delta_{i_{1}j_{2}}'K\left(\frac{\delta_{i_{1}j_{1}}^{2}}{c_{l_{1}}h_{n}}\right)K\left(\frac{\delta_{i_{1}j_{2}}^{2}}{c_{l_{2}}h_{n}}\right)\right] \\ 
&+ \frac{1}{r_{nl_{1}}c_{l_{1}}r_{nl_{2}}c_{l_{2}}h_{n}^{2}}E\left[\Delta_{i_{1}j_{1}}\Delta_{i_{2}j_{2}}'K'\left(\frac{\delta_{i_{1}j_{1}}^{2}}{c_{l_{1}}h_{n}}\right)K'\left(\frac{\delta_{i_{2}j_{2}}^{2}}{c_{l_{2}}h_{n}}\right)\left(F_{i_{1}j_{1}t_{1} s_{11}s_{12}} - \delta_{i_{1}j_{1}}^{2}\right)\left(F_{i_{2}j_{2}t_{1}s_{21}s_{22}} - \delta_{i_{2}j_{2}}^{2}\right)\right] \\
&+ \frac{4}{r_{nl_{1}}c_{l_{1}}r_{nl_{2}}c_{l_{2}}h_{n}^{2}}E\left[\Delta_{i_{1}j_{1}}\Delta_{i_{2}j_{2}}'K'\left(\frac{\delta_{i_{1}j_{1}}^{2}}{c_{l_{1}}h_{n}}\right)K'\left(\frac{\delta_{i_{2}j_{2}}^{2}}{c_{l_{2}}h_{n}}\right)\left(F_{i_{1}j_{1}t_{1}s_{11}s_{12}} - \delta_{i_{1}j_{1}}^{2}\right)\left(F_{i_{2}j_{2}t_{2}s_{11}s_{22}} - \delta_{i_{2}j_{2}}^{2}\right)\right]
\end{align*}
with $\Delta_{ij} = (x_{i}-x_{j})'(u_{i}-u_{j})$, $u_{i} = \lambda(w_{i}) + \varepsilon_{i}$, $\delta_{ij} = \delta(w_{i},w_{j}) = ||p_{w_{i}} - p_{w_{j}}||_{2}$, and $F_{ijts_{1}s_{2}} = f(w_{t},w_{s_{1}})f(w_{t},w_{s_{2}})\left(f(w_{i},w_{s_{1}})-f(w_{j},w_{s_{1}})\right)\left(f(w_{i},w_{s_{2}}) -f(w_{j},w_{s_{2}})\right)$.
\end{flushleft}


\subsubsection{Variance estimation}
Let $\hat{u}_{i} = y_{i} - x_{i}\hat{\beta}$,
\begin{align*}
\hat{\Gamma}_{h}  =  {n \choose 2}^{-1}\sum_{i=1}^{n-1}\sum_{j = i +1}^{n}\left(x_{i}-x_{j}\right)'\left(x_{i}-x_{j}\right)K\left(\frac{\hat{\delta}^{2}_{ij}}{h}\right)
\end{align*} 
and 
\begin{align*}
\hat{\Omega}_{n,h_{1}h_{2}} &= \frac{1}{n^{3}}\sum_{i = 1}^{n}\sum_{j_{1} = 1}^{n}\sum_{j_{2} = 1}^{n}\hat{\Delta}_{ij_{1}}\hat{\Delta}'_{ij_{2}}K\left(\frac{\hat{\delta}^{2}_{ij_{1}}}{h_{1}}\right)K\left(\frac{\hat{\delta}^{2}_{ij_{2}}}{h_{2}}\right) \\
&+ \frac{1}{n^{5}h_{{1}}h_{{2}}}\sum_{i_{1} = 1}^{n}\sum_{i_{2} = 1}^{n}\sum_{j_{1} = 1}^{n}\sum_{j_{2} = 1}^{n}\sum_{t = 1}^{n}\hat{\Delta}_{i_{1}j_{1}}\hat{\Delta}_{i_{2}j_{2}}K'\left(\frac{\hat{\delta}^{2}_{i_{1}j_{1}}}{h_{{1}}}\right)K'\left(\frac{\hat{\delta}^{2}_{i_{2}j_{2}}}{h_{2}}\right)\left(\hat{F}_{i_{1}j_{1}t} - \hat{\delta}_{i_{1}j_{1}}^{2}\right)\left(\hat{F}_{i_{2}j_{2}t} - \hat{\delta}_{i_{2}j_{2}}^{2}\right) \\
&+ \frac{4}{n^{5}h_{{1}}h_{{2}}}\sum_{i_{1} = 1}^{n}\sum_{i_{2} = 1}^{n}\sum_{j_{1} = 1}^{n}\sum_{j_{2} = 1}^{n}\sum_{t = 1}^{n}\hat{\Delta}_{i_{1}j_{1}}\hat{\Delta}_{i_{2}j_{2}}K'\left(\frac{\hat{\delta}^{2}_{i_{1}j_{1}}}{h_{{1}}}\right)K'\left(\frac{\hat{\delta}^{2}_{i_{2}j_{2}}}{h_{2}}\right)\left(\hat{F}'_{i_{1}j_{1}t} - \hat{\delta}_{i_{1}j_{1}}^{2}\right)\left(\hat{F}'_{i_{2}j_{2}t} - \hat{\delta}_{i_{2}j_{2}}^{2}\right)
\end{align*}
where $h_{l} = c_{l}h_{n}$, $\hat{\Delta}_{ij} = (x_{i}-x_{j})'(\hat{u}_{i} - \hat{u}_{j})$, $\hat{F}_{ijt} = \frac{1}{n^{2}}\sum_{s_{1}=1}^{n}\sum_{s_{2} = 1}^{n}D_{ts_{1}}D_{ts_{2}}\left(D_{is_{1}}-D_{js_{1}}\right)\left(D_{is_{2}}-D_{js_{2}}\right)$, and $\hat{F}'_{ijs_{1}} = \frac{1}{n^{2}}\sum_{t=1}^{n}\sum_{s_{2} = 1}^{n}D_{ts_{1}}D_{ts_{2}}\left(D_{is_{1}}-D_{js_{1}}\right)\left(D_{is_{2}}-D_{js_{2}}\right)$.
\begin{flushleft}
\textbf{Proposition C4}: Suppose Assumptions 1-4. Then $\left(\hat{\Gamma}_{h_{n}}^{-1} \hat{\Omega}_{n,h_{n},h_{n}}   \hat{\Gamma}_{h_n}^{-1} - nV_{4,n}\right) \to_{p} 0$ and $\left(\sum_{l_{1}=1}^{L}\sum_{l_{2}=1}^{L}a_{l_{1}}a_{l_{2}} \hat{\Gamma}_{c_{l_{1}}h_{n}}^{-1} \hat{\Omega}_{n,c_{l_{1}}h_{n},c_{l_{2}}h_{n}}   \hat{\Gamma}_{c_{l_{2}}h_{n}}^{-1} - nV_{5,n}\right) \to_{p} 0$ as $n \to \infty$.
\end{flushleft}

A corollary to Proposition C4 is that $\hat{\Gamma}_{h_{n}}^{-1} \hat{\Omega}_{h_{n},h_{n}}   \hat{\Gamma}_{h}^{-1}$ also consistently estimates $nV_{3,n}$ under the hypothesis of Proposition C1, although one can omit the last two summands when computing $\hat{\Omega}_{n,h_{n}h_{n}}$ because under the assumption that $f_{w_{i}}$ has finite support they are asymptotically negligible.

\subsection{Large sample properties of $\widehat{\lambda(w_{i})}$}
This section considers an estimator for $\lambda(w_{i})$ based on the bias-corrected estimator for $\beta$ from Section C.1.2, $\bar{\beta}_{L}$. That is, 
\begin{align*}
\overline{\lambda(w_{i})}_{L} = \left(\sum_{t=1}^{n}K\left(\frac{\hat{\delta}^{2}_{it}}{h_{n}}\right)\right)^{-1}\left(\sum_{t=1}^{n}\left(y_{t} - x_{t}\bar{\beta}_{L}\right)K\left(\frac{\hat{\delta}^{2}_{it}}{h_{n}}\right)\right).
\end{align*} 
Under additional restrictions on the choice of bandwidth sequence, the conditions in the hypothesis of Proposition 2 are also sufficient for $\{\overline{\lambda(w_{i})}_{L} \}_{i \in 
\mathcal{C}}$, the collection of estimators corresponding to a finite (i.e. fixed in $n$) set of agents $\mathcal{C} \subset \{1,...,n\}$, to be asymptotically normal. This additional restriction is given by Assumption C4 and the result is stated as the following Proposition C5. 

\begin{flushleft}
\textbf{Assumption C4:} The bandwidth sequence $h_{n}$ satisfies $\inf_{u \in [0,1]} b_{n}(u)n/r_{n}(u) \to 0$,  where $r_{n}(u) = E\left[K\left(\frac{||p_{u}-p_{w_j}||_{2}}{h_{n}}\right)\right]$,  $r_{n}'(u) = E\left[\lambda(w_{j})K\left(\frac{||p_{u}-p_{w_j}||_{2}}{h_{n}}\right)\right]$, and $b_{n}(u) = \left(\lambda(w_{i})r_{n}(u) - r_{n}'(u)\right)^{2}$. 
\end{flushleft}

The condition $\inf_{u \in [0,1]} b_{n}(u)n/r_{n}(u) \to 0$ is an undersmoothing condition that assumes that the bandwidth is chosen to be small enough so that the estimators are asymptotically unbiased. It can be approximated in practice using the empirical analogues of $r_{n,i}$, $r'_{n,i}$, and $\lambda(w_{i})$  \citep[see also the discussion after Assumption 4 in Section 2.3.4 of][]{auerbach2019identification}. 

\begin{flushleft} 
\textbf{Proposition C5}: Suppose Assumptions 1-4 and C1-4 hold. Let $\overline{\lambda}_{\mathcal{C}} = \{\overline{\lambda(w_{i})}_{L}\}_{i \in \mathcal{C}}$ for some finite collection of agents $\mathcal{C}$, $r_{n,i} :=  r_{n}(w_{i})$ and  $r_{n,i}' := r_{n}'(w_{i})$. Then as $n \to \infty$
\begin{align*}
V_{8,n}^{-1/2}\left(\bar{\lambda}_{\mathcal{C}} - \lambda_{\mathcal{C}}\right) \to_{d} \mathcal{N}\left(0,I_{|\mathcal{C}|}\right)
\end{align*}
where $\lambda_{\mathcal{C}} = \{\lambda(w_{i})\}_{i \in \mathcal{C}}$, $I_{|\mathcal{C}|}$ is the $|\mathcal{C}|\times|\mathcal{C}|$ identity matrix, and 
\begin{align*}
V_{8,n,ij} = \frac{1}{nr_{n,i}r_{n,j}}\sum_{t=1}^{n}&\left(\left(u_{t}K\left(\frac{\delta_{it}}{h_{n}}\right) - r_{n,i}'\right) - \frac{r_{n,i}'}{r_{n,i}}\left(K\left(\frac{\delta_{it}}{h_{n}}\right) - r_{n,i}\right)\right)\\
\times&\left(\left(u_{t}K\left(\frac{\delta_{jt}}{h_{n}}\right) - r_{n,j}'\right) - \frac{r_{n,j}'}{r_{n,j}}\left(K\left(\frac{\delta_{jt}}{h_{n}}\right) - r_{n,j}\right)\right).
\end{align*}
\end{flushleft}

One can estimate $V_{8,n,ij}$ directly using the empirical analogues of $u_{t}$, $\delta_{it}$, $r_{n,i}$ and $r_{n,i}'$, along the lines of Proposition C4. Consistency of the resulting variance estimators follows from identical arguments, and so is not demonstrated here. The problem of extending Proposition C5 to hold over all of $\{1,...,n\}$ is left to future work.

\subsection{Proof of results in Sections C.1 and C.2}
The proof of Proposition C1 uses the assumption that the link functions have finite support to strengthen Lemma B1 to the following Lemma C1.
\begin{flushleft} 
\textbf{Lemma C1}: Suppose Assumption 4 and $f_{w_{i}}$ has finite support. Then there exists an $\epsilon > 0$ such that $\max_{i\neq j} \hat{\delta}^{2}_{ij}\times\mathbbm{1}\{ \hat{\delta}^{2}_{ij} \leq \epsilon/2\} = o_{p}(n^{-1/2}h_{n})$.
\end{flushleft}

\begin{flushleft}
\textbf{Proof of Lemma C1}:
The assumption that $f_{w_{i}}$ has finite support implies there exists an $\epsilon > 0$ such that $\delta^{2}_{ij}1\{\delta^{2}_{ij} \leq \epsilon\} = 0$ and $\left(p_{w_{i}w_{t}}-p_{w_{j}w_{t}} \right) \times 1\{\delta^{2}_{ij} \leq \epsilon/2\} = 0$ for every $i,j \in \{1,...,n\}$ with probability one. For such an $\epsilon$, write
\begin{align*}
\hat{\delta}^{2}_{ij}1\{\hat{\delta}^{2}_{ij} \leq \epsilon/2\} = \hat{\delta}^{2}_{ij}\left(1\{\hat{\delta}^{2}_{ij} \leq \epsilon/2\} - 1\{\delta^{2}_{ij} \leq \epsilon/2\}\right) + \hat{\delta}^{2}_{ij}1\{\delta^{2}_{ij} \leq \epsilon/2\}.
\end{align*}
I show that both summands are $o_{p}\left(n^{-1/2}h_{n}\right)$. First, $\max_{i \neq j}\sqrt{n}h_{n}^{-1}\hat{\delta}^{2}_{ij}1\{\delta^{2}_{ij} \leq \epsilon/2\} = o_{p}(1)$ because $\left(p_{w_{i}w_{t}}-p_{w_{j}w_{t}} \right) \times 1\{\delta^{2}_{ij} \leq \epsilon/2\} = 0$, Bernstein's inequality, and the union bound imply
\begin{align*}
P\left(\max_{i,j,t}\left[(n-3)^{-1}\sum_{s \neq i,j,t}D_{ts}(D_{is}-D_{js})\right]^{2}1\{\delta^{2}_{ij} \leq \epsilon/2\} \geq \eta\right) \leq 2n^{3}\exp\left(\frac{-(n-3)\eta}{3}\right)
\end{align*} 
and averaging over $t \neq i,j$ gives
\begin{align*}
P\left(\max_{i,j}\sqrt{n}h_{n}^{-1}\hat{\delta}^{2}_{ij}1\{\delta^{2}_{ij} \leq \epsilon/2\} \geq \eta\right) \leq 2n^{3}\exp\left(\frac{-(n-3)\eta h_{n}}{3\sqrt{n}}\right)  = o(1).
\end{align*}

Second, since $\delta^{2}_{ij} \in (\epsilon/4, 3\epsilon/4)$ is a probability zero event and $\hat{\delta}_{ij}^{2} \leq 1$ for every $i,j \in \{1,...,n\}$,
\begin{align*}
\sqrt{n}h_{n}^{-1}|\hat{\delta}^{2}_{ij}\left(1\{\hat{\delta}^{2}_{ij} \leq \epsilon/2\} - 1\{\delta^{2}_{ij} \leq \epsilon/2\}\right)| &\leq 2\sqrt{n}h_{n}^{-1}\times 1\{|\hat{\delta}^{2}_{ij} - \delta^{2}_{ij}| > |\epsilon/2 - \delta^{2}_{ij}|\} \\
&\leq  2\sqrt{n}h_{n}^{-1}1\{|\hat{\delta}^{2}_{ij} - \delta^{2}_{ij}| > \epsilon/4\}
\end{align*}
and so $\max_{i\neq j}\sqrt{n}h_{n}^{-1}|\hat{\delta}^{2}_{ij}\left(1\{\hat{\delta}^{2}_{ij} \leq \epsilon/2\} - 1\{\delta^{2}_{ij} \leq \epsilon/2\}\right)| = o_{p}(1)$ by Lemma B1. $\square$ \newline
\end{flushleft}


\begin{flushleft}
\textbf{Proof of Proposition C1}:
In the proof of Proposition 2, I demonstrate that Assumptions 1-4 are sufficient for
\begin{align*}
\frac{1}{m} \sum_{i}\sum_{j >i }(x_{i}-x_{j})'(x_{i}-x_{j})K\left(\frac{\hat{\delta}_{ij}^{2}}{h_{n}}\right) \to_{p} \Gamma_{0}E\left[K\left(\frac{\delta_{ij}^{2}}{h_{n}}\right)\right]
\end{align*}
where $m = n(n-1)/2$ and $\delta_{ij} = \delta(w_{i},w_{j})$. Since the support of $f_{w_{i}}$ is finite, $E\left[K\left(\frac{\delta_{ij}^{2}}{h_{n}}\right)\right]$ $= K(0)P(\delta_{ij} = 0) > 0$ eventually.
\newline

As for the numerator, I follow the proof of Proposition 2 and write
\begin{align*}
U_{n} = \frac{1}{m} \sum_{i}\sum_{j > i}\Delta_{ij}K\left(\frac{\hat{\delta}^{2}_{ij}}{h_{n}}  \right) = \frac{1}{m} \sum_{i}\sum_{j > i} \Delta_{ij}\left[K\left(\frac{\delta^{2}_{ij}}{h_{n}}\right) + K'\left(\frac{\iota_{ij}}{h_{n}}\right)\left(\frac{\hat{\delta}^{2}_{ij} - \delta^{2}_{ij}}{h_{n}}\right)1\{\hat{\delta}^{2}_{ij} \leq h_{n}\}  \right] 
\end{align*}
where $\iota_{ij}$ is a mean value between $\delta^{2}_{ij}$ and $\hat{\delta}^{2}_{ij}$ and $\Delta_{ij} = \left(x_{i}-x_{j}\right)'\left(u_{i}-u_{j}\right)$. I first show that $\frac{1}{m} \sum_{i}\sum_{j > i} \Delta_{ijl}K'\left(\frac{\iota_{ij}}{h_{n}}\right)\left(\frac{\hat{\delta}^{2}_{ij} - \delta_{ij}^{2}}{h_{n}}\right)1\{\hat{\delta}^{2}_{ij} \leq h_{n}\} = o_{p}\left(n^{-1/2}\right)$ where $\Delta_{ijl}$ is the $l$th component of $\Delta_{ij}$. By Cauchy-Schwartz
\begin{align*}
\frac{1}{m} &\left|\sum_{i}\sum_{j > i} \left(\Delta_{ijl}K'\left(\frac{\iota_{ij}}{h_{n}}\right)\left(\frac{\hat{\delta}^{2}_{ij} - \delta^{2}_{ij}}{h_{n}}\right)  \right)\right|\\
& \leq \frac{\bar{K}'}{m}\left(\sum_{i}\sum_{j > i}\Delta_{ijl}^{2}\right)^{1/2}\times \left(\sum_{i}\sum_{j > i}\left(\frac{\hat{\delta}^{2}_{ij} - \delta^{2}_{ij}}{h_{n}}\right)^{2}1\{\hat{\delta}^{2}_{ij} \leq h_{n}\}\right)^{1/2}
\end{align*}
where $\bar{K}' = \sup_{u \in [0,1]}K'(u)$, $\sum_{i}\sum_{j > i}\Delta_{ijl}^{2} = O_{p}(m)$ since $x_{i}$ and $u_{i}$ have finite fourth moments, and $\max_{i \neq j}\left(\frac{\hat{\delta}^{2}_{ij} - \delta^{2}_{ij}}{h_{n}}\right)1\{\hat{\delta}^{2}_{ij} \leq h_{n}\} = o_{p}\left(n^{-1/2}\right)$ by Lemma C1. 
\newline

It follows that
\begin{align*}
U_{n} = \frac{1}{m}\sum_{i}\sum_{j > i}\Delta_{ij}K\left(\frac{\delta_{ij}^{2}}{h_{n}}\right)  +  o_{p}\left(n^{-1/2}\right).
\end{align*}
The first summand is a second order U-statistic with symmetric $L^{2}$-integrable kernel, so by Lemma A.3 of \cite{ahn1993semiparametric}
\begin{align*}
\sqrt{n}\left(U_{n} - U\right) \to \mathcal{N}(0,V)
\end{align*}
where $U = E\left[\Delta_{ij}K\left(\frac{\delta^{2}_{ij}}{h_{n}}\right)\right] $ and for $Z_{i} = (x_{i},\varepsilon_{i},w_{i})$
\begin{align*}
V = \lim_{h \to 0}4 E\left[E\left[\Delta_{ij}K\left(\frac{\delta^{2}_{ij}}{h}\right)\hspace{1mm} | \hspace{1mm} Z_{i}\right] E\left[\Delta'_{ij}K\left(\frac{\delta^{2}_{ij}}{h}\right)\hspace{1mm} | \hspace{1mm} Z_{i}\right] \right] = \lim_{h \to 0}4 E\left[\Delta_{ij}\Delta_{ik}'K\left(\frac{\delta^{2}_{ij}}{h}\right)K\left(\frac{\delta^{2}_{ik}}{h}\right) \right].
\end{align*}
Since $f_{w_{i}}$ has finite support, $E[\delta_{ij}^{2}|\delta_{ij}^{2} \leq \epsilon]= 0$ for some $\epsilon > 0$, and so $U = E\left[\Delta_{ij}K\left(0\right)1\{\delta_{ij} = 0\}\right] = 0$ for $n$ sufficiently large such that $h_{n} \leq \epsilon$. Similarly $V= 4\Omega_{0}K(0)^{2}P\left(\delta_{ij}=0, \delta_{ij} = 0\right)$. So by Slutsky's Theorem, 
\begin{align*}
\sqrt{n}\left(\hat{\beta} - \beta \right) \to_{d} \mathcal{N}(0,V_{3})
\end{align*}
where $V_{3} =  \Gamma_{0}^{-1}\Omega_{0}\Gamma_{0}^{-1} \times s$ as claimed. $\square$ \newline
\end{flushleft}

\begin{flushleft}
\textbf{Proof of Proposition C2}:
The proof of Proposition 2 demonstrates that Assumptions 1-3 and C2 are sufficient for the denominator to converge in probability to $\Gamma_{n}$. As for the numerator,
\begin{align*}
&U_{n} = \frac{1}{{n \choose 2}r_{n}} \sum_{i}\sum_{j > i} \Delta_{ij}K\left(\frac{\hat{\delta}_{ij}^{2}}{h_{n}}\right) \\
&= \frac{1}{{n \choose 2}r_{n}}\sum_{i}\sum_{j > i}\Delta_{ij}\left[K\left(\frac{\delta_{ij}^{2}}{h_{n}}\right) + K'\left(\frac{\delta_{ij}^{2}}{h_{n}}\right)\left(\frac{\hat{\delta}_{ij}^{2} - \delta_{ij}^{2}}{h_{n}}\right) + K''\left(\frac{\iota_{ij}}{h_{n}}\right)\left(\frac{\hat{\delta}_{ij}^{2} - \delta_{ij}^{2}}{h_{n}}\right)^{2} \right]
\end{align*}
where $\iota_{ij}$ is the intermediate value between $\hat{\delta}_{ij}^{2}$ and $\delta_{ij}^{2}$ suggested by Taylor and the mean value theorem. 
\newline

First, I show that
\begin{align*}
\frac{1}{{n \choose 2}r_{n}}\sum_{i}\sum_{j > i} \Delta_{ij}K''\left(\frac{\iota_{ij}}{h_{n}}\right)\left(\frac{\hat{\delta}_{ij}^{2} - \delta_{ij}^{2}}{h_{n}}\right)^{2} = o_{p}\left(n^{-1/2}\right)
\end{align*}
Let $s_{n} = n^{-1/2}h_{n}^{2}r_{n}$. Since $\{(u,v): \delta_{uv} \leq \varepsilon\} \geq \left(\frac{\varepsilon}{C}\right)^{1/\alpha}$ by the first part of Lemma 1 and Assumption C1, $r_{n} \geq \underline{K}C^{-1/\alpha}h_{n}^{1/\alpha}$ for $\underline{K}  = \liminf_{h \to 0}E\left[K\left(\frac{\delta_{ij}}{h} \right)|\delta_{ij} \leq h\right] > 0$ by choice of kernel density function $K$ in Assumption 4. Since $n^{1/2-\gamma}h_{n}^{2+1/\alpha} \to \infty$ for some $\gamma > 0$ by choice of $h_{n}$ in Assumption C2, $n^{1-\gamma}s_{n} \to \infty$, and so Lemma B1 implies that $\max_{i \neq j}\left(\frac{\hat{\delta}_{ij}^{2} - \delta_{ij}^{2}}{\sqrt{s_{n}}}\right)^{2} = o_{p}(1)$ or $\max_{i \neq j}\left(\frac{\hat{\delta}_{ij}^{2} - \delta_{ij}^{2}}{h_{n}\sqrt{r_{n}}}\right)^{2} = o_{p}\left(n^{-1/2}\right)$. It follows that
\begin{align*}
\frac{1}{{n \choose 2}r_{n}}\sum_{i}&\sum_{j > i}\Delta_{ij}K''\left(\frac{\iota_{ij}}{h_{n}}\right)\left(\frac{\hat{\delta}_{ij}^{2} - \delta_{ij}^{2}}{h_{n}}\right)^{2} \leq \frac{\bar{K}''}{{n \choose 2}}\sum_{i}\sum_{j > i}\Delta_{ij} \times o_{p}\left(n^{-1/2}\right)
\end{align*}
where $\bar{K}'' = \sup_{u \in [0,1]}K''(u)$ and the last line is $o_{p}\left(n^{-1/2}\right)$ because $x_{i}$ and $u_{i}$ have finite fourth moments. It follows from this first step that
\begin{align*}
U_{n} = \frac{1}{{n \choose 2}r_{n}}\sum_{i}\sum_{j > i} \Delta_{ij}\left[K\left(\frac{\delta_{ij}^{2}}{h_{n}}\right) + K'\left(\frac{\delta_{ij}^{2}}{h_{n}}\right)\left(\frac{\hat{\delta}_{ij}^{2} - \delta_{ij}^{2}}{h_{n}}\right) \right] + o_{p}\left(n^{-1/2}\right).
\end{align*}
\newline
Second, I show that
\begin{align*}
U_{n} = \frac{1}{{n \choose 5}^{2}r_{n}}\sum_{i}\sum_{j > i}\sum_{t > j}\sum_{s_{1} >t}\sum_{s_{2} > s_{1}} \Delta_{ij}\left[K\left(\frac{\delta_{ij}^{2}}{h_{n}}\right) + \frac{1}{h_{n}} K'\left(\frac{\delta_{ij}^{2}}{h_{n}}\right)\left(F_{ijts_{1}s_{2}} - \delta_{ij}^{2}\right)\right]  +  o_{p}\left(n^{-1/2}\right)
\end{align*}
where $F_{ijts_{1}s_{2}} = f_{ts_{1}}f_{ts_{2}}(f_{is_{1}}-f_{js_{1}})(f_{is_{2}}-f_{js_{2}})$. Let $\tilde{\delta}^{2}_{ij} = {n - j\choose 3}^{-1}\sum_{t > j}\sum_{s_{1} > t}\sum_{s_{2} > s_{1}}F_{ijts_{1}s_{2}}$. Then
\begin{align*}
U_{n} &= \frac{1}{{n \choose 2}r_{n}}\sum_{i}\sum_{j > i} \Delta_{ij}\left[K\left(\frac{\delta_{ij}^{2}}{h_{n}}\right) + K'\left(\frac{\delta_{ij}^{2}}{h_{n}}\right)\left(\frac{\tilde{\delta}_{ij}^{2} - \delta_{ij}^{2}}{h_{n}}\right) \right]\\
&+ \frac{1}{{n \choose 2}r_{n}}\sum_{i}\sum_{j > i} \Delta_{ij}\left[K'\left(\frac{\delta_{ij}^{2}}{h_{n}}\right)\left(\frac{\hat{\delta}_{ij}^{2} - \tilde{\delta}_{ij}^{2}}{h_{n}}\right) \right] + o_{p}\left(n^{-1/2}\right)
\end{align*}
and the second summand is $o_{p}(n^{-1/2})$ by Chebyshev's inequality, since it has mean zero and variance
\begin{align*}
\frac{1}{{n \choose 2} ^{2}n^{6}r_{n}^{2}h_{n}^{2}}E\left[\sum_{i_{1}}\sum_{i_{2}}\sum_{j_{1}}\sum_{j_{2}} \sum_{t_{1}}\sum_{t_{2}}\sum_{s_{11}}\sum_{s_{12}} \sum_{s_{21}}\sum_{s_{22}} \Delta_{i_{1}j_{1}}\Delta_{i_{2}j_{2}}'K'\left(\frac{\delta_{i_{1}j_{1}}^{2}}{h_{n}}\right)K'\left(\frac{\delta_{i_{2}j_{2}}^{2}}{h_{n}}\right) \right. \\
\times\left(D_{i_{1}j_{1}t_{1}s_{11}s_{12}} - F_{i_{1}j_{1}t_{1}s_{11}s_{12}} \right)\times\left( D_{i_{2}j_{2}t_{2}s_{21}s_{22}} - F_{i_{2}j_{2}t_{2}s_{21}s_{22}}  \right) \Bigg]
\end{align*}
where $D_{ijts_{1}s_{2}} =  D_{ts_{1}}D_{ts_{2}}(D_{is_{1}}-D_{js_{1}})(D_{is_{2}}-D_{js_{2}})$. To see that this variance is $o\left(n^{-1}\right)$, note that unless two elements from the set $\{i_{1},j_{1},t_{1},s_{11},s_{12}\}$ equal two in $\{i_{2},j_{2},t_{2},s_{21},s_{22}\}$, $\{\eta_{t_{1}s_{11}}, \eta_{t_{1}s_{12}}, \eta_{i_{1}s_{11}}, \eta_{j_{1}s_{11}}, \eta_{i_{1}s_{12}}, \eta_{j_{1}s_{12}}\}$ is independent of  $\{\eta_{t_{2}s_{21}}, \eta_{t_{2}s_{22}}, \eta_{i_{2}s_{21}}, \eta_{j_{2}s_{21}}, \eta_{i_{2}s_{22}}, \eta_{j_{2}s_{22}}\}$ and so
\begin{align*}
 E \Bigg[\left[ D_{i_{1}j_{1}t_{1}s_{11}s_{12}} - F_{i_{1}j_{1}t_{1}s_{11}s_{12}} \right]\times\left[ D_{i_{2}j_{2}t_{2}s_{21}s_{22}} - F_{i_{2}j_{2}t_{2}s_{21}s_{22}}\right]
 |  Z_{i_{1}j_{1}t_{1}s_{11}s_{12}}, Z_{i_{2}j_{2}t_{2}s_{21}s_{22}}  \Bigg] = 0
\end{align*}
where $Z_{i} = \{x_{i},w_{i},\nu_{i}\}$ and $Z_{ijts_{1}s_{2}} = \{Z_{i},Z_{j},Z_{t},Z_{s_{1}},Z_{s_{2}}\}$. Since $K'\left(\frac{\delta_{i_{1}j_{1}}^{2}}{h}\right)$ is $O_{p}(r_{n})$, $nh_{n}^{4} \to \infty$ implies that this variance is $o\left(n^{-1}\right)$ and so the second summand is $o_{p}\left(n^{-1/2}\right)$ . 
\newline

Let
\begin{align*}
U_{n}' &= \frac{1}{{n \choose 5}^{2}r_{n}}\sum_{i}\sum_{j > i}\sum_{t > j}\sum_{s_{1} >t}\sum_{s_{2} > s_{1}} \Delta_{ij}\left[K\left(\frac{\delta_{ij}^{2}}{h_{n}}\right) + \frac{1}{h_{n}} K'\left(\frac{\delta_{ij}^{2}}{h_{n}}\right)\left(F_{ijts_{1}s_{2}} - \delta_{ij}^{2}\right)\right]   \\
&= U_{n} +  o_{p}\left(n^{-1/2}\right).
\end{align*}
$U_{n}'$ is a 5th order U-statistic with kernel function depending on the sample size. It can be represented by the following iid sum \citep[see Lemma 3.2 of][]{powell1989semiparametric}
\begin{align*}
U_{n} &= E[U'_{n}] +\frac{2}{nr_{n}}\sum_{\tau =1}^{n} \left(E\left[\Delta_{\tau j}K\left(\frac{\delta_{\tau j}^{2}}{h_{n}}\right)|Z_{\tau}\right] - E[U'_{n}]\right) \\
&+ \frac{1}{nr_{n}h_{n}}\sum_{\tau=1}^{n}E\left[\Delta_{ij}K'\left(\frac{\delta_{ij}^{2}}{h_{n}}\right)\left(F_{ij\tau s_{1}s_{2}} - \delta_{ij}^{2}\right)|Z_{\tau} \right]\\
 &+ \frac{2}{nr_{n}h_{n}}\sum_{\tau=1}^{n}E\left[\Delta_{ij}K'\left(\frac{\delta_{ij}^{2}}{h_{n}}\right)\left(F_{ijt\tau s_{2}} - \delta_{ij}^{2}\right)|Z_{\tau} \right] + o_{p}\left(n^{-1/2}\right)
\end{align*}
where $E[U'_{n}] = r_{n}^{-1}E\left[\Delta_{ij}K\left(\frac{\delta_{ij}^{2}}{h_{n}}\right)\right]$ and $Z_{\tau} = \{ x_{\tau},w_{\tau}, \nu_{\tau}\}$. In particular, $U_{n}$ can be represented asymptotically by an iid sum of random variables, so by the Lindeberg Central Limit Theorem
\begin{align*}
\Omega_{n}^{-1/2}\left(U_{n} - E[U_{n}]\right) \to_{d} \mathcal{N}\left(0,I_{k}\right)
\end{align*}
where for a collection of ten distinct agents $\{i_{1},i_{2},j_{1},j_{2},t_{1},t_{2},s_{11},s_{12},s_{21}s,_{22}\}$
\begin{align*}
\Omega_{n} &= \frac{4}{r_{n}^{2}}E\left[\Delta_{i_{1}j_{1}}\Delta_{i_{1}j_{2}}'K\left(\frac{\delta_{i_{1}j_{1}}^{2}}{h_{n}}\right)K\left(\frac{\delta_{i_{2}j_{2}}^{2}}{h_{n}}\right)\right] \\ 
&+ \frac{1}{r_{n}^{2}h_{n}^{2}}E\left[\Delta_{i_{1}j_{1}}\Delta_{i_{2}j_{2}}'K'\left(\frac{\delta_{i_{1}j_{1}}^{2}}{h_{n}}\right)K'\left(\frac{\delta_{i_{2}j_{2}}^{2}}{h_{n}}\right)\left(F_{i_{1}j_{1}t_{1} s_{11}s_{12}} - \delta_{i_{1}j_{1}}^{2}\right)\left(F_{i_{2}j_{2}t_{1}s_{21}s_{22}} - \delta_{i_{2}j_{2}}^{2}\right)\right] \\
&+ \frac{4}{r_{n}^{2}h_{n}^{2}}E\left[\Delta_{i_{1}j_{1}}\Delta_{i_{2}j_{2}}'K'\left(\frac{\delta_{i_{1}j_{1}}^{2}}{h_{n}}\right)K'\left(\frac{\delta_{i_{2}j_{2}}^{2}}{h_{n}}\right)\left(F_{i_{1}j_{1}t_{1}s_{11}s_{12}} - \delta_{i_{1}j_{1}}^{2}\right)\left(F_{i_{2}j_{2}t_{2}s_{11}s_{22}} - \delta_{i_{2}j_{2}}^{2}\right)\right]
\end{align*}
since $E[U'_{n}] \to_{p} 0$ by Proposition 2. It follows from Slutsky's Theorem that
\begin{align*}
V_{4,n}^{-1/2}\left(\hat{\beta} - \beta - \left(\Gamma_{n}\right)^{-1}E\left[U'_{n}\right] \right) \to_{d} \mathcal{N}\left(0,I_{k} \right)
\end{align*}
which demonstrates the claim. $\square$ \newline
\end{flushleft}

\begin{flushleft}
\textbf{Proof of Proposition C3}:
Since $\bar{\beta}_{L} = \sum_{l=1}^{L}a_{l}\hat{\beta}_{C_{l}h_{n}}$, a trivial extension of the proof of Proposition C2 to the collection of estimators $\{\hat{\beta}_{C_{l}h_{n}}\}_{l=1}^{L}$ using the continuous mapping theorem is
\begin{align*}
V_{5,n}^{-1/2}\left(\bar{\beta}_{L} - \bar{\beta}_{L,h_{n}}\right) =  V_{5,n}^{-1/2}\sum_{l=1}^{L}a_{l}\left(\hat{\beta}_{C_{l}h_{n}} - \beta_{C_{l}h_{n}}\right) \to_{d} \mathcal{N}\left(0,I_{k}\right)
\end{align*}
where $\bar{\beta}_{L,h} = \sum_{l=1}^{L}a_{l}\beta_{C_{l}h}$ is the pseudo-truth associated with $\beta_{L}$. By Assumption C3, the estimator can also be written as
\begin{align*}
\bar{\beta}_{L,h} &= \beta + \sum_{l_{1} =1}^{L}\sum_{l_{2} =1}^{L}a_{l_{1}}\left(2\Gamma_{0}\right)^{-1}C_{l_{2}}\left(c_{l_{1}}h \right)^{l_{2}/\theta} + o_{p}\left(n^{-1/2}\right) \\
&= \beta + \left(2\Gamma_{0}\right)^{-1}\sum_{l_{2}}C_{l_{2}} \left[\sum_{l_{1}}a_{l_{1}}c_{l_{1}}^{l_{2}/\theta}\right]h^{l_{2}/\theta}  + o_{p}\left(n^{-1/2}\right)
\end{align*}
since $\sum_{l_{2}}a_{l_{2}} = 1$ by choice of $\{a_{1},...,a_{L}\}$. The second summand is $0$ because, $\{a_{1},...,a_{L}\}$ also satisfies $\left[\sum_{l_{1}}a_{l_{1}}c_{l_{1}}^{l_{2}/\theta}\right] = 0$ for all $l_{2} \in \{1,...,L\}$. The claim follows. $\square$
\newline
\end{flushleft}


\begin{flushleft}
\textbf{Proof of Proposition C4:} 
I demonstrate the second claim, which includes the first as a special case. The proof of Proposisition 2 demonstrates that Assumptions 1-4 are sufficient for $r_{n,c}^{-1}\hat{\Gamma}_{ch_{n}} = \Gamma_{n} + o_{p}(1)$ for any $c > 0$ where $\delta_{ij} = \delta(w_{i},w_{j})$ and $r_{n,c} = \left(E\left[K
\left(\frac{\delta_{ij}}{ch_{n}}\right)\right]\right)$. It remains to be shown that $\left(r_{n,c_{1}}r_{n,c_{2}}\right)^{-1}\hat{\Omega}_{c_{1}h_{n},c_{2}h_{n}}$ converges to $\Omega_{nc_{1}c_{2}}$ for any $c_{1}, c_{2} > 0$. I consider the three terms that make up $\hat{\Omega}_{c_{1}h_{n},c_{2}h_{n}}$ seperately.
\newline

The first term is $\frac{1}{n^{3}r_{n,c_{1}}r_{n,c_{2}}}\sum_{i = 1}^{n}\sum_{j_{1} = 1}^{n}\sum_{j_{2} = 1}^{n}\hat{\Delta}_{ij_{1}}\hat{\Delta}_{ij_{2}}K\left(\frac{\hat{\delta}_{ij_{1}}}{h_{1}}\right)K\left(\frac{\hat{\delta}_{ij_{2}}}{h_{2}}\right)$, where $\hat{\Delta}_{ij} = (x_{i}-x_{j})'(\hat{u}_{i}-\hat{u}_{j})$ and $\hat{u}_{i} = y_{i} - x_{i}\hat{\beta}$. Lemma A1 and Proposition 2 imply that $\hat{\delta}_{ij} = \delta_{ij} + o(1)$ and $\hat{u}_{i} = u_{i} + o(1)$ where $u_{i} = y_{i} - x_{i}\beta$, and so the term converges to $\frac{1}{n^{3}r_{n,c_{1}}r_{n,c_{2}}}\sum_{i = 1}^{n}\sum_{j_{1} = 1}^{n}\sum_{j_{2} = 1}^{n}\Delta_{ij_{1}}\Delta'_{ij_{2}}K\left(\frac{\delta_{ij_{1}}}{h_{1}}\right)K\left(\frac{\delta_{ij_{2}}}{h_{2}}\right)$ by the continuous mapping theorem, which is a third order V-statistic in the sense of \cite{ahn1993semiparametric}, and thus converges in probability to $\frac{1}{r_{n,c_{1}}r_{n,c_{2}}}E\left[\Delta_{ij_{1}}\Delta'_{ij_{2}}K\left(\frac{\delta_{ij_{1}}}{h_{1}}\right)K\left(\frac{\delta_{ij_{2}}}{h_{2}}\right)\right]$. 
\newline

The second term is \[\frac{1}{n^{5}c_{1}h_{{1}}r_{n,c_{1}}c_{2}h_{{2}}r_{n,c_{2}}}\sum_{i_{1} = 1}^{n}\sum_{i_{2} = 1}^{n}\sum_{j_{1} = 1}^{n}\sum_{j_{2} = 1}^{n}\sum_{t = 1}^{n}\hat{\Delta}_{i_{1}j_{1}}\hat{\Delta}_{i_{2}j_{2}}K'\left(\frac{\hat{\delta}^{2}_{i_{1}j_{1}}}{c_{1}h_{{1}}}\right)K'\left(\frac{\hat{\delta}^{2}_{i_{2}j_{2}}}{c_{2}h_{2}}\right)\left(\hat{F}_{i_{1}j_{1}t} - \hat{\delta}_{i_{1}j_{1}}^{2}\right)\left(\hat{F}_{i_{2}j_{2}t} - \hat{\delta}_{i_{2}j_{2}}^{2}\right)\]
where $\hat{F}_{ijt} = \frac{1}{n^{2}}\sum_{s_{1}=1}^{n}\sum_{s_{2} = 1}^{n}D_{ts_{1}}D_{ts_{2}}\left(D_{is_{1}}-D_{js_{1}}\right)\left(D_{is_{2}}-D_{js_{2}}\right)$. By previous arguments this converges to the fifth-order V-statistic
\[\frac{1}{n^{5}c_{1}h_{{1}}r_{n,c_{1}}c_{2}h_{{2}}r_{n,c_{2}}}\sum_{i_{1} = 1}^{n}\sum_{i_{2} = 1}^{n}\sum_{j_{1} = 1}^{n}\sum_{j_{2} = 1}^{n}\sum_{t = 1}^{n}\Delta_{i_{1}j_{1}}\Delta_{i_{2}j_{2}}K'\left(\frac{\delta^{2}_{i_{1}j_{1}}}{c_{1}h_{{1}}}\right)K'\left(\frac{\delta^{2}_{i_{2}j_{2}}}{c_{2}h_{2}}\right)\left(F_{i_{1}j_{1}t} - \delta_{i_{1}j_{1}}^{2}\right)\left(F_{i_{2}j_{2}t} - \delta_{i_{2}j_{2}}^{2}\right)\]
where $F_{ijt} = E\left[D_{ts_{1}}D_{ts_{2}}\left(D_{is_{1}}-D_{js_{1}}\right)\left(D_{is_{2}}-D_{js_{2}}\right)|w_{i},w_{j},w_{t}\right]$ is the probability limit of $\hat{F}_{ijt}$. The second term thus converges to
\[\frac{1}{c_{1}h_{n}r_{n,c_{1}}c_{2}h_{2}r_{n,c_{2}}}E\left[\Delta_{i_{1}j_{1}}\Delta_{i_{2}j_{2}}K'\left(\frac{\delta^{2}_{i_{1}j_{1}}}{c_{1}h_{{1}}}\right)K'\left(\frac{\delta^{2}_{i_{2}j_{2}}}{c_{2}h_{2}}\right)\left(F_{i_{1}j_{1}t} - \delta_{i_{1}j_{1}}^{2}\right)\left(F_{i_{2}j_{2}t} - \delta_{i_{2}j_{2}}^{2}\right)\right].\]
\newline

The third term is \[\frac{4}{n^{5}h_{{1}}h_{{2}}}\sum_{i_{1} = 1}^{n}\sum_{i_{2} = 1}^{n}\sum_{j_{1} = 1}^{n}\sum_{j_{2} = 1}^{n}\sum_{t = 1}^{n}\hat{\Delta}_{i_{1}j_{1}}\hat{\Delta}_{i_{2}j_{2}}K'\left(\frac{\hat{\delta}_{i_{1}j_{1}}}{h_{{1}}}\right)K'\left(\frac{\hat{\delta}_{i_{2}j_{2}}}{h_{2}}\right)\left(\hat{F}'_{i_{1}j_{1}t} - \hat{\delta}_{i_{1}j_{1}}^{2}\right)\left(\hat{F}'_{i_{2}j_{2}t} - \hat{\delta}_{i_{2}j_{2}}^{2}\right)\]
where $\hat{F}'_{ijs_{1}} = \frac{1}{n^{2}}\sum_{t=1}^{n}\sum_{s_{2} = 1}^{n}D_{ts_{1}}D_{ts_{2}}\left(D_{is_{1}}-D_{js_{1}}\right)\left(D_{is_{2}}-D_{js_{2}}\right)$. By previous arguments this converges to the fifth order V-statistic
\[\frac{4}{n^{5}h_{{1}}h_{{2}}}\sum_{i_{1} = 1}^{n}\sum_{i_{2} = 1}^{n}\sum_{j_{1} = 1}^{n}\sum_{j_{2} = 1}^{n}\sum_{t = 1}^{n}\Delta_{i_{1}j_{1}}\Delta_{i_{2}j_{2}}K'\left(\frac{\delta_{i_{1}j_{1}}}{h_{{1}}}\right)K'\left(\frac{\delta_{i_{2}j_{2}}}{h_{2}}\right)\left(F'_{i_{1}j_{1}t} - \delta_{i_{1}j_{1}}^{2}\right)\left(F'_{i_{2}j_{2}t} - \delta_{i_{2}j_{2}}^{2}\right)\]
where $F'_{ijs_{1}} = E\left[D_{ts_{1}}D_{ts_{2}}\left(D_{is_{1}}-D_{js_{1}}\right)\left(D_{is_{2}}-D_{js_{2}}\right)|w_{i},w_{j},w_{s_{1}}\right]$ is the probability limit of $\hat{F}'_{ijs_{1}}$. The third term thus converges to
\[\frac{1}{c_{1}h_{n}r_{n,c_{1}}c_{2}h_{2}r_{n,c_{2}}}E\left[\Delta_{i_{1}j_{1}}\Delta_{i_{2}j_{2}}K'\left(\frac{\delta_{i_{1}j_{1}}}{h_{{1}}}\right)K'\left(\frac{\delta_{i_{2}j_{2}}}{h_{2}}\right)\left(F'_{i_{1}j_{1}t} - \delta_{i_{1}j_{1}}^{2}\right)\left(F'_{i_{2}j_{2}t} - \delta_{i_{2}j_{2}}^{2}\right)\right].\]

The claim then follows from the continuous mapping theorem. $\square$
\end{flushleft}

\begin{flushleft}
\textbf{Proof of Proposition C5:}
The proof of Proposition C5 closely follows that of Propositions 2 and C2, and so only a sketch is provided here. Let $\lambda_{i}$, $\bar{\lambda}_{i}$, and $\delta_{it}$ shorthand $\lambda(w_{i})$, $\overline{\lambda(w_{i})_{L}}$, and $\delta(w_{i},w_{t})$ respectively. Then 
\begin{align*}
\left(\bar{\lambda}_{i}r_{n,i} - r_{n,i}'\right) &= \frac{1}{n}\sum_{t=1}^{n}\left(\bar{u}_{t}K\left(\frac{\hat{\delta}_{it}}{h_{n}}\right) - r_{n,i}'\right) - \frac{r_{n,i}'}{r_{n,i}}\left(K\left(\frac{\hat{\delta}_{it}}{h_{n}}\right) - r_{n,i}\right) + rem_{n,i} \\
&= \frac{1}{n}\sum_{t=1}^{n}\left(\left[u_{t} - x_{t}\left(\bar{\beta}_{L} - \beta\right)\right]K\left(\frac{\delta_{it}}{h_{n}}\right) - r_{n,i}'\right) - \frac{r_{n,i}'}{r_{n,i}}\left(K\left(\frac{\delta_{it}}{h_{n}}\right) - r_{n,i}\right) + o_{p}\left(\sqrt{\frac{r_{n,i}}{n}}\right) + rem_{n,i}\\
&= \frac{1}{n}\sum_{t=1}^{n}\left(u_{t}K\left(\frac{\delta_{it}}{h_{n}}\right) - r_{n,i}'\right) - \frac{r_{n,i}'}{r_{n,i}}\left(K\left(\frac{\delta_{it}}{h_{n}}\right) - r_{n,i}\right) + o_{p}\left(\sqrt{\frac{r_{n,i}}{n}}\right) + rem_{n,i}
\end{align*}
where $\bar{u}_{t} = y_{t} - x_{t}\bar{\beta}_{L}$, $r_{n,i} = E\left[K\left(\frac{\delta_{it}}{h_{n}}\right)|w_{i}\right]$, $r_{n,i}' = E\left[u_{t}K\left(\frac{\delta_{it}}{h_{n}}\right)|w_{i}\right]$, and $rem_{i,n}$ is an error that is stochastically small relative to the first two summands (it is the remainder from a first order Taylor approximation). The second equality follows from the fact that $\max_{i\neq j}|\hat{\delta}_{ij}-\delta_{ij}|$ is $o_{p}\left(\left(nr_{n,i}\right)^{-1/2}\right)$ from Lemma B1 (just take $\gamma$ close to $1$). The third equality follows from the fact that $||\bar{\beta}_{L}-\beta||$ is  $o_{p}\left(\left(nr_{n,i}\right)^{-1/2}\right)$ from Proposition C3.  
\newline

Let  $\theta_{n} = \frac{1}{n}\sum_{t=1}^{n}\left\{\left(u_{t}K\left(\frac{\delta_{it}}{h_{n}}\right) - r_{n,i}'\right) - \frac{r_{n,i}'}{r_{n,i}}\left(K\left(\frac{\delta_{it}}{h_{n}}\right) - r_{n,i}\right)\right\}_{i \in \mathcal{C}}$. Since the right-hand side sum has independent entries with bounded third moments, it follows from the Lindeberg Central Limit Theorem that $V_{8,n}^{-1/2}\theta_{n} \to_{d} \mathcal{N}\left(0,I_{|\mathcal{C}|}\right)$ where the $ij$th entry of $V_{8,n}$ is given by 
\begin{align*} 
V_{8,n,ij} = n^{-1}\sum_{t=1}^{n}\left(\left(u_{t}K\left(\frac{\delta_{it}}{h_{n}}\right) - r_{n,i}'\right) - \frac{r_{n,i}'}{r_{n,i}}\left(K\left(\frac{\delta_{it}}{h_{n}}\right) - r_{n,i}\right)\right)\left(\left(u_{t}K\left(\frac{\delta_{jt}}{h_{n}}\right) - r_{n,j}'\right) - \frac{r_{n,j}'}{r_{n,j}}\left(K\left(\frac{\delta_{jt}}{h_{n}}\right) - r_{n,j}\right)\right).
\end{align*}
Since $b_{n,i}n/r_{n,i} \to_{p} 0$ for all $i \in \mathcal{C}$, $V_{8,n,ij}^{-1/2}r_{n,i}\left(\theta - \bar{\lambda}_{i,L}\right) = o_{p}(1)$. The claim follows. $\square$ 
\end{flushleft}

\section{Extensions}
This section sketches how some of the assumptions of Section 2 of \cite{auerbach2019identification} might be relaxed. Section 2.1 considers weighted and directed networks, Section 2.2 considers models with multiple social characteristics, Section 2.3 considers link covariates, and Section 2.4 considers models with endogenous social characteristics. 

\subsection{Weighted and directed networks}
Weighted and directed networks may be represented by adjacency matrices with entries that take values in $\mathbb{R}^{L}$ for some positive integer $L$. For example, $D_{ij} = (D_{ij}[1],D_{ij}[2]) \in \mathbb{R}^{2}$ might record the amount of trade between countries $i$ and $j$ where the first element $D_{ij}[1]$ is the amount of exports from $i$ to $j$ and the second element $D_{ij}[2]$ is the amount of exports from $j$ to $i$. A model for this case is 
\begin{align*}
\mathbbm{1}\{D_{ij} \leq_{L} x\} = \mathbbm{1}\{\eta_{ij} \leq f(w_{i},w_{j};x)\}.
\end{align*}
where $\leq_{L}$ refers to some partial order on $\mathbb{R}^{L}$. In this case, heterogeneity in linking behavior can be characterized by the weighted link function $f(w_{i},\cdot;\cdot): [0,1]\times \mathbb{R}^{L} \to [0,1]$ and differences between two link functions can be measured by the network distance
\begin{align*}
d(w_{i},w_{j}) =  \max_{x \in \mathbb{R}^{L}}\left(\int \left(f(w_{i},\tau;x)-f(w_{j},\tau;x)\right)^{2}d\tau \right)^{1/2}.
\end{align*}
Since this model is equivalent to (2) for any fixed $x$, the arguments of Section 2 can be directly applied. That is, one can instead consider conditional expectations defined with respect to the maximum codegree distance
\begin{align*}
\delta(w_{i},w_{j}) =  \max_{x \in \mathbb{R}^{L}}\left(\int\left( \int \left(f(w_{i},s;x)-f(w_{j},s;x)\right)f(\tau,s;x)\right)^{2}d\tau\right)^{1/2}
\end{align*}
and construct estimators based on its empirical analog 
\begin{align*}
\hat{\delta}_{ij} =  \max_{x \in \mathbb{R}^{L}}\left(\frac{1}{n}\sum_{t=1}^{n}\left(\frac{1}{n}\sum_{s=1}^{n}\mathbbm{1}\{D_{ts} \leq x\}\left(\mathbbm{1}\{D_{is} \leq x\}-\mathbbm{1}\{D_{js} \leq x\}\right)\right)^{2}\right)^{1/2}
\end{align*}
where $\hat{\delta}_{ij}$ can be computed in practice by taking the maximization over the $n \choose 2$ unique elements of the adjacency matrix $D$.

\subsection{Multiple social characteristics}
The restriction that the social characteristics $\{w_{i}\}_{i=1}^{n}$ are real-valued is non-essential. In fact, if the social characteristics are supported on a compact subset $R$ of $\mathbb{R}^{L}$  for some positive integer $L$, Propositions 1 and 2 in Section 2 still hold \emph{mutatis mutandis}. The main idea of this paper is to use the linking functions $\{f_{w_{i}}\}_{i=1}^{n}$ to incorporate unobserved heterogeneity into the regression model, and this is always a square-integrable function, regardless of the dimension of $w_{i}$. If $w_{i}$ is high-dimensional, then the size of the support of $f_{w_{i}}$ may be relatively large, so that the proposed estimators converge at a relatively slow rate. Unless the researcher has access to additional information about the network formation process, this is an unavoidable consequence of working with high-dimensional network data at the level of generality of (2) in \cite{auerbach2019identification}.

\subsection{Link covariates}
In many settings, the probability that two agents form a link is thought to depend on observed covariates and the researcher would like to control for the variation in linking behavior induced by these covariates. I refer to these variables as link covariates to distinguish them from those in the right-hand side of the regression model (1). An application of link covariates is to identify network peer effects. See Online Appendix Section E.1. 

Link covariates can be incorporated into the framework of Section 2 by considering the network formation model
\begin{align*}
D_{ij} = \mathbbm{1}\{\eta_{ij} \leq f_{z}(w_{i},w_{j}; z_{ij})\}
\end{align*}  
where $z_{ij} = z(w_{i},w_{j},\xi_{ij}) \in \mathbb{R}^{L}$ denote the link covariates and $\{\xi_{ij}\}_{i \neq j}$ are independent and identically distributed with standard uniform marginals that are mutually independent of $\{w_{i}\}_{i=1}^{n}$ and $\{\eta_{ij}\}_{i,j=1}^{n}$. For instance, $z_{ij}$ may indicate whether students $i$ and $j$  take a class together or participate in the same extracurricular activity. The network formation model of this subsection is can also be represented by the unconditional model $D_{ij} = \mathbbm{1}\{\eta_{ij} \leq f(w_{i},w_{j})\}$ where $f(w_{i},w_{j}) = E\left[ f_{z}(w_{i},w_{j}; z_{ij})|w_{i},w_{j}\right]$ and thus cast as a special case of the network formation model from Section 2 (without link covariates)

To incorporate the link covariates, I propose characterizing agent $i$'s linking behavior using the conditional link function $f_{w_{i};z} = f_{z}(w_{i},\cdot;\cdot):[0,1]\times\mathcal{Z}\to\mathbb{R}$ and conditional network distance 
 \begin{align*}
 d_{z}(w_{i},w_{j}) =  \left(\int\int \left(f_{z}(w_{i},\tau;\zeta)- f_{z}(w_{j},\tau;\zeta)\right)^{2} d\tau d\zeta\right)^{1/2}.
 \end{align*}
 This is as opposed to the unconditional link function $f(w_{i},\cdot) =E\left[ f_{z}(w_{i},w_{j}; z_{ij})|w_{i},w_{j}= \cdot\right]$ and unconditional network distance 
 \begin{align*}
 d(w_{i},w_{j}) =  \left(\int \left(E\left[ f_{z}(w_{i},w_{j}; z_{ij})|w_{i},w_{j}= \tau\right]- E\left[ f_{z}(w_{i},w_{j}; z_{ij})|w_{i},w_{j}= \tau\right]\right)^{2} d\tau\right)^{1/2}
 \end{align*}
 that characterizes linking behavior in the network if the link covariates are ignored. 

I also propose estimation by conditional codegree matching. Let the conditional codegree function refer to $p_{z}(w_{i},\cdot) = \int\int f_{z}(w_{i},s;\zeta)f_{z}(\cdot,s;\zeta)g(w_{i},s,\zeta)dsd\zeta$ and the conditional codegree distance between agents $i$ and $j$ refer to  
\begin{align*}
\delta_{z}(w_{i},w_{j}) = \left(\int\left(\int\int\left(f_{z}(\tau,s;\zeta)\left(f_{z}(w_{i},s;\zeta)- f_{z}(w_{j},s;\zeta)\right)\right)g(\tau,s;\zeta)dsd\zeta\right)^{2}d\tau\right)^{1/2}. 
\end{align*}
where $g(\tau,s;\zeta)$ denotes the conditional density of $z_{ij}$ given $(w_{i},w_{j}) = (\tau,s)$ evaluated at $\zeta$. 

Conditional codegree distance has an empirical analog
\begin{align*}
\hat{\delta}_{ij;z}^{2} = \frac{1}{n}\sum_{t=1}^{n}\left[\frac{\sum_{s=1}^{n}D_{ts}(D_{is}-D_{js})K_{z}\left(\frac{\left|z_{is} - z_{ts}\right|+ |z_{js} - z_{ts}|}{h_{n;z}}\right)}{\sum_{s=1}^{n}K_{z}\left(\frac{\left|z_{is} - z_{ts}\right|+ |z_{js} - z_{ts}|}{h_{n;z}}\right)}\right]^{2}
\end{align*}
where $K_{z}$ is a kernel density function and $h_{n;z}$ is a bandwidth sequence satisfying certain regularity conditions (see Online Appendix Section E.1.4 below). The estimator is premised on the following covariate overlap condition.
\begin{flushleft}
\textbf{Assumption D1:} The conditional distribution of $\left(z_{is},z_{js}\right)$ given $(w_{i},w_{j})$ is smooth and square-integrable, with full support on $\mathcal{Z}\times\mathcal{Z}$. The function $f$ satisfies the continuity condition that $\inf_{u \in [0,1]}\int \mathbbm{1}\{v \in [0,1], z \in \mathcal{Z}: \sup_{\tau \in [0,1]} |f(u,\tau:z) - f(v,\tau;z)| \leq \varepsilon\} > 0$.
\end{flushleft}

Intuitively, the first condition supposes that for two students $i$ and $j$ there exist other students that take classes with both $i$ and $j$, don't take classes with both $i$ and $j$, and take classes with one but not the other. This could be the case when there is independent variation that drives the link covariates independent of the agent social characteristics. The second condition is an implication of continuity of $f$ \citep[see Section 2.1 of][]{auerbach2019identification}. 
%

\subsection{Endogeneity}
The assumption in Section 2 that $E[\varepsilon_{i}|w_{i},x_{i}] = 0$, can also be relaxed. The social characteristics may be endogenous when they are determined by an omitted variable that also drives variation in the outcome that is different from the social influence of interest. For example, suppose that $y_{i}$ is student GPA, $x_{i}$ is an indicator for whether agent $i$ participates in a tutoring program, and $w_{i}$ is a positive integer that indexes the social clique to which student $i$ belongs. Suppose that $\varepsilon_{i} = \mu(r_{i}) + \nu_{i}$ where $r_{i} \in \mathbb{R}^{L}$ measures student $i$'s socioeconomic status, $\nu_{i}$ is an idiosyncratic error independent of $(w_{i},x_{i},r_{i})$, and $\mu$ is an unknown smooth function. The endogeneity problem is that $E[\varepsilon_{i}|w_{i},x_{i}] \neq 0$ because a student's decision to enroll in the tutoring program, join a specific social clique, and their baseline GPA are all potentially related to that student's socioeconomic status. In other words, $E\left[\mu(r_{i})|w_{i},x_{i}\right] \neq 0$.

The parameters of interest are the impact of the tutoring program $\beta$ and the baseline GPA of agent $i$'s social clique $\lambda(w_{i})$. If the researcher can measure socioeconomic status or a proxy, then they can add an additional control function into the right-hand side of the regression model. That is, write
\begin{align*}
y_{i} = x_{i}\beta + \lambda(w_{i}) + \mu(r_{i}) + \nu_{i}.
\end{align*}
Following the logic of Section 2, $\beta$ and $\lambda(w_{i})$ are identified (the latter up to a constant) if, for example, Assumption 2 is strengthened to 
\begin{align*}
E\left[\left(x_{i}-x_{j}\right)'\left(x_{i}-x_{j}\right)|\hspace{1mm}||f_{w_i}-f_{w_j}||_{2} = 0, ||r_{i}-r_{j}||_{2} = 0\right]
\end{align*}
 is positive definite and $r_{i}$ does not perfectly predict $f_{w_{i}}$. Estimators can be constructed along the lines of Section 2.3. 

If the researcher cannot measure socioeconomic status directly, then another potential option is to use instrument variables. Suppose, for example, that the researcher observes data on the students' class schedules, where $z_{i} \in \mathbb{R}^{L}$ indicates a number of possible class schedules in which agent $i$ could be enrolled.  Class assignment may be unrelated to socioeconomic status if it is determined arbitrarily by school officials so that $E\left[\varepsilon_{i}|z_{i}\right] = 0$. The parameters of the regression model are then identified so long as classroom assignment drives sufficient variation in social clique formation and information about the tutoring program. For example, one could replace Assumption 2 with 
\begin{align*}
E\left[\left(E[x_{i}|z_{i}]-E[x_{j}|z_{j}]\right)'\left(E[x_{i}|z_{i}]-E[x_{j}|z_{j}]\right)|\hspace{1mm}||f_{w_i}-f_{w_j}||_{2} = 0\right] \text{ is positive definite }
\end{align*}
and Assumption 3 with
\begin{align*}
E\left[E\left[\lambda(w_{i})|z_{i}\right]-E\left[\lambda(w_{j})|z_{j}\right] |\hspace{1mm}||f_{w_i}-f_{w_j}||_{2} = 0\right] = 0.
\end{align*}
Intuitively, the first assumption states that there is variation in expected program participation amongst students in the same social clique but across different class schedules. This may be the case, for example, if the program was advertised in some classes but not others. The second assumption states that the students in a given class schedule all have the same baseline GPA due to social clique. This may be the case, for example, if the student class schedule drives all of the variation in clique formation that is relevant to determine the social influence. 

The identification conditions suggest a three step estimator for $\beta$ and $\lambda(w_{i})$. In the first step, the researcher projects $y_{i}$ and $x_{i}$ onto $z_{i}$. Denote the predicted values by $y^{z}_{i} = \hat{E}\left[y_{i}|z_{i}\right]$ and $x^{z}_{i} = \hat{E}\left[x_{i}|z_{i}\right]$. In the second step, the researcher estimates $\beta$ via the pairwise difference estimator of (4), except replacing $y_{i}$ and $x_{i}$ with $y_{i}^{z}$ and $x_{i}^{z}$. In the third step, the researcher estimates $\lambda(w_{i})$ using the nonparametric regression (5), except using the residuals $y_{i}^{z}-x_{i}^{z}\hat{\beta}^{z}$ where $\hat{\beta}^{z}$ is the estimator from the second step. 

\section{Network peer effects}
This section considers an application to a network version of the \cite{manski1993identification} linear-in-means peer effects model. The model is related to previous work by  \cite{bramoulle2009identification, de2010identification, goldsmith2013social, hsieh2014social, johnsson2015estimation,arduini2015parametric}, and others.
\subsection{Model}
 Let $y_{i}$ be student GPA, $x_{i}$ be a vector of student covariates (age, grade, gender, etc.), and $D_{ij} = 1$ if students $i$ and $j$ are peers and $0$ otherwise. One extension of the \cite*{manski1993identification} linear-in-means peer effects model to the network setting is
\begin{align}
y_{i} &= x_{i}\beta + E[x_{j}|D_{ij} = 1,w_{i}]\rho_{1} + \rho_{2}E[y_{j}|D_{ij} = 1,w_{i}]  + \lambda(w_{i}) + \varepsilon_{i} \label{lim} \\
D_{ij} &= \mathbbm{1}\{ \eta_{ij} \leq f(w_{i},w_{j})\}\mathbbm{1}\{i \neq j\} \nonumber
\end{align}
where $w_{i}$ measures student $i$'s social ability \citep[or student ambition as motivated by][]{jackson2014networks}, $E[x_{j}|D_{ij} = 1,w_{i}]$ denotes the expected covariates of student $i$'s peers, $E[y_{j}|D_{ij} = 1,w_{i}]$ denotes the expected GPA of student $i$'s peers, and $\lambda(w_{i})$ is the direct effect of social ability on GPA (for instance, students with more social ability might pay more attention in class).\footnote{ I follow the convention of defining $E\left[x_{j}|D_{ij}=1,w_{i}\right] = E\left[y_{j}|D_{ij}=1,w_{i}\right]  = 0$ whenever $P(D_{ij}=1|w_{i}) = 0$. In other words, agents with no peers have no peer influence.} In Manski's terminology, the parameters $(\rho_{1},\rho_{2},\lambda(w_{i}))$ refer to exogenous, endogenous, and correlated peer effects respectively. Identification problems stemming from the fact that all three terms are functions of $w_{i}$ is discussed below. \cite{bramoulle2009identification,de2010identification,goldsmith2013social, chan2014additive, hsieh2014social, johnsson2015estimation} and \cite{arduini2015parametric} consider related  models with additional restrictions on $\lambda$ or $f$.

The use of the expected peer outcomes $E[y_{j}|D_{ij} = 1,w_{i}]$ instead of averages $\sum_{j=1}^{n}y_{j}D_{ij}/\sum_{j=1}^{n}D_{ij}$ reflects a particular interpretation about the model and sampling procedure: the peer groups that determine student GPA are not exactly given in the random sample drawn by the researcher. In contrast, the above literature typically assumes that the researcher has sampled all of the other agents (students, teachers, family members, etc.) whose outcomes and characteristics influence student $i$'s GPA. See \cite{manski1993identification} for a discussion.

The model (\ref{lim}) is not the only extension of the liner-in-means peer effects model to the network setting. Another potential model is 
\begin{align}\label{otherlim}
y_{i} &=   x_{i}\beta + E[x_{i}|w_{i}]\rho_{1} + \rho_{2}E[y_{i}|w_{i}]  + \lambda(w_{i}) + \varepsilon_{i} \\
D_{ij} &= \mathbbm{1}\{ \eta_{ij} \leq f(w_{i},w_{j})\}\mathbbm{1}\{i \neq j\} \nonumber,
\end{align}
where $E[x_{i}|w_{i}]$ is the expected covariates of student $i$ and $E[y_{i}|w_{i}]$ is the expected GPA of student $i$ given that student's social ability. The model differs from (\ref{lim}) in that students react to their own expected outcomes and covariates rather than the expected outcomes and covariates of their peers. While the two models are identical in the group peer effects setting, they can have very different implications in the network setting. For example, the parameters of (\ref{otherlim}) are not generally identified because of the reflection problem. The parameters of (\ref{lim}) are (under certain conditions) identified, as demonstrated by \cite{bramoulle2009identification}.
 
\subsection{Interpretation}
Since \cite{goldsmith2013social} and \cite{jackson2014networks}, the literature on social interactions has typically viewed (2) as a literal description of how an exogenous attribute, such as social ability, informs agent linking decisions. The quantity $ f(w_{i},w_{j})-\eta_{ij}$ is interpreted as a random utility that agents $i$ and $j$ receive from forming a link. Under this interpretation, the model (2) restricts the types of preferences agents can have over the configuration of social connections realized in the population. 

Probably the most contentious assumption is conditional independence: the utility agents $i$ and $j$ receive from forming a link only depends on the linking decisions between agents $k$ and $l$ through their social characteristics $\{w_{i},w_{j},w_{k},w_{l}\}$. This assumption is also made by \cite{holland1981exponential, duijn2004p2, krivitsky2009representing, mccormick2012latent, dzemski2014empirical, jochmans2016semiparametric, nadler2016networked,  graham2017econometric, candelaria2016semiparametric, gao2017nonparametric,toth2017semiparametric} \citep[see generally][Sections 3 and 6]{graham2019network}, but violated in settings where some link formation is endogenous \citep[see for instance][]{sheng2012identification, leung2015two, ridder2015estimation, menzel2015strategic, mele2017approximate}. That is, agents' incentives to form links explicitly depend on which other links are realized and the researcher observes an equilibrium in which no linked pair of agents would prefer to destroy their link and no unlinked pair of agents would prefer to create a link \citep[see recently][]{jackson1996strategic}. 

When some link formation is endogenous, the network formation model (2) can still be viewed as a reduced-form approximation to the within-equilibrium distribution of network links \citep[see generally][Section 3.4]{graham2019network}. However, two agents with similar such collections of link probabilities may have very different social characteristics, so that Assumption 3 is potentially false, and the main identification strategy of the paper is potentially invalid. Recently \cite{griffith2016random} and \cite{badev2017discrete} consider parametric models of social interaction and network formation that, among other things, explicitly account for endogenous link formation. The extent to which the parametric structure in their models can be relaxed as in (2) is, to my knowledge, an open question. A recent step in this direction is provided by \cite{mele2017structural}. With some additional information endogenous link formation is testable, see recent work by \cite{auerbach2019measuring,pelican2020optimal} 

\subsection{Identification}
If $\lambda$ is a constant function, then the parameters $(\beta,\rho_{1},\rho_{2})$ of equation (\ref{lim}) are identified if $\{c_{i},x_{i},E[x_{j}|D_{ij} = 1,w_{i}],E[y_{j}|D_{ij} = 1,w_{i}]\}_{i=1}^{n}$ are linearly independent where $c$ is a constant vector. 

When $\lambda$ is not constant, the parameter $\beta$ is identified provided Assumptions 2-3 hold by Proposition 1. The parameters $\rho_{1}$ and $\rho_{2}$ are not separately identified from $\lambda$ without more information because
\begin{align*}
E[x_{j}|D_{ij} = 1,w_{i}] = E[x_{j}D_{ij}|w_{i}]/E[D_{ij}|w_{i}]  = \int E[x_{j}|w_{j} = w]f(w_{i},w)dw/\int f(w_{i},w)dw
\end{align*}
 is a continuous functional of $f_{w_{i}}$. Non-identification stems from the fact that under (2), the link function $f_{w_{i}}$ determines agent $i$'s expected peer group and thus the expected covariates and outcomes of that peer group. Under Assumptions 2-3, $\lambda(w_{i})$ is also determined by the link function $f_{w_{i}}$. Consequently, holding $f_{w_{i}}$ fixed, there is no residual variation in agent $i$'s expected peer group that can be used to separately identify $\rho_{1}$ and $\rho_{2}$ from $\lambda(w_{i})$. 

Identification of $(\rho_{1},\rho_{2})$ is restored so long as the researcher is willing to impose an additional restriction that distinguishes either the expected peer covariates and outcomes or the unknown social influence term from an generic functional of $f_{w_{i}}$. One way to do this is to incorporate additional observed covariates that drive either link formation or define the peer groups, but not both. I provide two examples. 

In the first example, the expected peer outcomes and covariates depend on other agent-specific variables. For instance, one could consider the model
\begin{align*}
y_{i} &= x_{i}\beta + E[x_{j}|D_{ij} = 1,w_{i},z_{i} = z_{j}]\rho_{1} + E[y_{j}|D_{ij} = 1,w_{i},z_{i} = z_{j}]\rho_{2}  + \lambda(w_{i}) + \varepsilon_{i} \\
D_{ij} &= \mathbbm{1}\{ \eta_{ij} \leq f(w_{i},w_{j})\}\mathbbm{1}\{i \neq j\}.
\end{align*}
where $z_{i}$ measures student age (assumed to be discrete or have density function bounded away from $0$), students are only influenced by the expected covariates and outcomes of similarly aged peers (although they report connections to other students of every age), the conditional distribution of social ability given student age has full support, and $E\left[\varepsilon_{i}|x_{i},w_{i},z_{i}\right] = 0$. The parameters of the regression model are then identified so long as there is residual variation in $(x_{i},E[x_{j}|D_{ij} = 1,w_{i},z_{i} = z_{j}], E[y_{j}|D_{ij} = 1,w_{i},z_{i} = z_{j}])$ not explained by $f_{w_{i}}$ and the social influence $\lambda(w_{i})$ is determined by $f_{w_{i}}$. This follows directly from Proposition 1 in Section 2.  

In the second example, the link decision rule depends on agent-pair-specific link covariates. This corresponds to the network formation model of Section 3.4, that is
\begin{align*}
y_{i} &= x_{i}\beta + E[x_{j}|D_{ij} = 1,w_{i}]\rho_{1} + E[y_{j}|D_{ij} = 1,w_{i}]\rho_{2}  + \lambda(w_{i}) + \varepsilon_{i} \\
D_{ij} &= \mathbbm{1}\{ \eta_{ij} \leq f_{z}(w_{i},w_{j};z_{ij})\}\mathbbm{1}\{i \neq j\},
\end{align*}
where $z_{ij} = z(w_{i},w_{j};\xi_{ij})$ measures whether students $i$ and $j$ are enrolled in the same class or participate in the same extracurricular activity, and $f_{w_{i;z}}(\cdot;\cdot) = f_{z}(w_{i},\cdot;\cdot)$. If $\lambda(w_{i}) = E\left[\lambda(w_{i})|f_{w_{i};z}\right]$ then there may be residual variation in $E[x_{j}|D_{ij} = 1,w_{i}]$ and $E[y_{j}|D_{ij} = 1,w_{i}]$ to identify $\rho_{1}$ and $\rho_{2}$ so long as there are agents with similar conditional link functions $f_{w_{i};z}$ but different unconditional link function $f_{w_{i}}(\cdot) = E\left[f_{z}(w_{i},w_{j};z_{ij})|w_{i},w_{j} = \cdot\right]$. This is because it is the latter function that determines the expected peer outcomes and covariates in this example. That is,
\begin{align*}
E[x_{j}|D_{ij} = 1,w_{i}] &= E[x_{j}D_{ij}|w_{i}]/E[D_{ij}|w_{i}]  \\ &= \int E[x_{j}|w_{j} = w]E\left[f_{z}(w_{i},w_{j};z_{ij})|w_{i},w_{j} = w\right]dw/\int E\left[f_{z}(w_{i},w_{j};z_{ij})|w_{i},w_{j} = w\right]dw.
\end{align*}

This intuition motivates the following identification Assumptions E2 and E3. Let $X_{i} \in \mathbb{R}^{2k+1}$ such that $X_{i}  = \left(x_{i},E[x_{j}|D_{ij} = 1,w_{i}],E[y_{j}|D_{ij} = 1,w_{i}]\right)$. The conditions are
\begin{flushleft}
\textbf{Assumption E2:} $\inf_{u \in [0,1]}\sigma_{2k+ 1}\left(E\left[\left(X_{i}-E\left[X_{i}|w_{i} = u\right]\right)'\left(X_{i}-E\left[X_{i}|w_{i} = u\right]\right)|\hspace{1mm} w_{i} = u\right]\right) > 0$ where $\sigma_{2k+1}(\cdot)$  is the smallest eigenvalue.
\end{flushleft}
and
\begin{flushleft}
\textbf{Assumption E3:} For every $\epsilon > 0$ there exists a $\delta > 0$ such that $\sup_{u,v \in [0,1]: ||f_{w_{i};z} - f_{u;z}||_{2} \leq \delta}\left(\lambda(u)  - \lambda(v)\right)^{2} \leq \epsilon$.
\end{flushleft}

The premise of Assumption E3 is that it is not the unconditional link function that determines the social influence in the regression model but the conditional link function. This may be the case, for example, if the part of social ability that determines classroom participation and directly influences GPA is unrelated to the part of social ability that determines classroom enrollment or participation in extracurricular activities. Only conditional on student class or activity does the network links produce relevant information about classroom participation to control for it and identify the social effects of interest. 

Let $\theta = (\beta, \rho_{1},\rho_{2})$. 
\begin{flushleft}
\textbf{Proposition E1}: Suppose Assumptions E2-E3. Then
\begin{itemize}
\item[(i)]  $\theta$ uniquely minimizes $E\left[\left(y_{i}-y_{j} - (X_{i}-X_{j})b\right)^{2}|\hspace{1mm}||f_{w_i;z}-f_{w_j;z}||_{2} = 0\right]$ over $b \in \mathbb{R}^{2k+1}$.
\item[(ii)] $\lambda(w_{i}) = E\left[\left(y_{i} - X_{i}\theta\right) |\hspace{1mm}f_{w_{i};z}\right]$.
\end{itemize}
\end{flushleft} 
The logic of Proposition 3 is identical to that of Proposition 1 and its proof can be found in Online Appendix Section E.5. Intuitively, to see how the link covariates restore identification suppose that two students who participate in the same extracurricular activity are more likely to form a link. Then if different extracurricular activities are associated with different distributions of social ability $w_{i}$ (but have overlapping support), then two students that have the same social ability (and thus would have the same peer groups if placed in the same extra-curricular activity) but participate in different activities may have different expected peer outcomes and covariates. It is this across-activity variation that identifies $\rho_{1}$ and $\rho_{2}$.

%

\subsection{Estimation}
First suppose that the researcher has access to agent-level variables that shift the expected peer outcomes and covariates as in the first example of Section 4.1.3. I propose the following estimators. For each $i = 1,...,n$ define $\hat{X}_{i} \in \mathbb{R}^{2k+1}$ such that $\hat{X}_{i} = (x_{i},0,...,0)$ if $\sum_{j=1}^{n}D_{ij}K_{z}\left(\frac{z_{i}-z_{j}}{h_{n;z}}\right) = 0$ and 
\begin{align*}
\hat{X}_{i} = \left(x_{i}, \frac{\sum_{j=1}^{n}x_{j}D_{ij}K_{z}\left(\frac{z_{i}-z_{j}}{h_{n;z}}\right)}{\sum_{j=1}^{n}D_{ij}K_{z}\left(\frac{z_{i}-z_{j}}{h_{n;z}}\right)}, \frac{\sum_{j=1}^{n}y_{j}D_{ij}K_{z}\left(\frac{z_{i}-z_{j}}{h_{n;z}}\right)}{\sum_{j=1}^{n}D_{ij}K_{z}\left(\frac{z_{i}-z_{j}}{h_{n;z}}\right)}\right)
\end{align*}
otherwise, where $K_{z}$ is a kernel density function and $h_{n;z}$ is a bandwidth parameter. Under additional regularity conditions, $\theta$ is consistently estimated by the pairwise difference estimator
 \begin{align*}
\hat{\theta} = &\left( \sum_{i=1}^{n-1}\sum_{j = i+1}^{n}(\hat{X}_{i}-\hat{X}_{j})'(\hat{X}_{i}-\hat{X}_{j})K\left(\frac{\hat{\delta}^{2}_{ij}}{h_{n}}\right)\right)^{-1} \left(\sum_{i=1}^{n-1}\sum_{j = i+1}^{n}(\hat{X}_{i}-\hat{X}_{j})'(y_{i}-y_{j})K\left(\frac{\hat{\delta}^{2}_{ij}}{h_{n}}\right)\right)
\end{align*}
where $\hat{\delta}_{ij}$ is the empirical codegree distance (3) and $\lambda(w_{i})$ is estimated using the residual variation
\begin{align*}
\widehat{\lambda(w_{i})} = \left(\sum_{t=1}^{n}K\left(\frac{\hat{\delta}^{2}_{it}}{h_{n}}\right)\right)^{-1}\left(\sum_{t=1}^{n}\left(y_{t} - \hat{X}_{t}\hat{\theta}\right)K\left(\frac{\hat{\delta}^{2}_{it}}{h_{n}}\right)\right).
\end{align*}
Consistency of $\hat{\theta}$ and $\widehat{\lambda(w_{i})}$ follows directly from Proposition 2 in Section 2 so long as $K_{z}$ and $h_{n;z}$ are chosen such that the average peer covariates and outcomes converge uniformly to their population analogues (i.e. $\max_{i = 1,...,n}|\hat{X}_{i} -X_{i}| = o_{p}(1)$). This is the case under standard regularity conditions \cite*[see for instance][]{powell1994estimation}. Accurate inference using standard tools likely requires trimming, see for instance \cite{robinson1988root}.  

Now suppose that the researcher has access to link covariates as in the second example of Section 4.1.3 and that Assumptions 5-7 are satisfied. I propose the following estimators. For each $i = 1,...,,n$ define $\hat{X}_{i} \in \mathbb{R}^{2k+1}$ such that $\hat{X}_{i} = (x_{i},0,...,0)$ if $\sum_{j=1}^{n}D_{ij} = 0$ and $\hat{X}_{i} = \left(x_{i}, \frac{\sum_{j=1}^{n}x_{j}D_{ij}}{\sum_{j=1}^{n}D_{ij}}, \frac{\sum_{j=1}^{n}y_{j}D_{ij}}{\sum_{j=1}^{n}D_{ij}}\right)$ otherwise. Under additional regularity conditions $\theta$ is consistently estimated by the pairwise difference estimator
 \begin{align*}
\hat{\theta} = &\left( \sum_{i=1}^{n-1}\sum_{j = i+1}^{n}(\hat{X}_{i}-\hat{X}_{j})'(\hat{X}_{i}-\hat{X}_{j})K\left(\frac{\hat{\delta}^{2}_{ij;z}}{h_{n}}\right)\right)^{-1} \left(\sum_{i=1}^{n-1}\sum_{j = i+1}^{n}(\hat{X}_{i}-\hat{X}_{j})'(y_{i}-y_{j})K\left(\frac{\hat{\delta}^{2}_{ij;z}}{h_{n}}\right)\right)
\end{align*}
where 
\begin{align*}
\hat{\delta}_{ij;z}^{2} = \frac{1}{n}\sum_{t=1}^{n}\left[\frac{\sum_{s=1}^{n}(D_{is}-D_{js})D_{ts}K_{z}\left(\frac{\left|z_{is} - z_{ts}\right|+ |z_{js} - z_{ts}|}{h_{n;z}}\right)}{\sum_{s=1}^{n}K_{z}\left(\frac{\left|z_{is} - z_{ts}\right|+ |z_{js} - z_{ts}|}{h_{n;z}}\right)}\right]^{2}
\end{align*}
and $\lambda(w_{i})$ is consistently estimated using the residual variation
\begin{align*}
\widehat{\lambda(w_{i})} = \left(\sum_{t=1}^{n}K\left(\frac{\hat{\delta}^{2}_{it;z}}{h_{n}}\right)\right)^{-1}\left(\sum_{t=1}^{n}\left(y_{t} - \hat{X}_{t}\hat{\theta}\right)K\left(\frac{\hat{\delta}^{2}_{it;z}}{h_{n}}\right)\right).
\end{align*}

The additional regularity conditions are on the kernel density function $K_{z}$ and bandwidth sequence $h_{n;z}$ used in the conditional codegree distance. This is given in the following Assumption E4.
\begin{flushleft}
\textbf{Assumption E4:} The bandwidth sequence satisfies $h_{n;z} \to 0$, $n^{1-\gamma}h_{n;z}^{2L+1} \to \infty$ for some $\gamma > 0$. $K_{z}$ is supported, bounded, and smooth on $[0,1)$. 
\end{flushleft}

The content of Assumption E4 is typical of the semiparametrics literature. It is sufficient for the following result. 
 \begin{flushleft} 
\textbf{Proposition E2}: Suppose Assumptions D1 and E2-E4 hold. Then $\left(\hat{\theta} - \theta\right) \to_{p} 0$ and $\max_{i = 1,...,n}\left|\widehat{\lambda(w_{i})} - \lambda(w_{i})\right| \to_{p} 0$ as $n \to \infty$ .
\end{flushleft}
The main difference in the proof of Proposition 4 and that of Proposition 2 is the extension of Lemma 1 it allow for link covariates.

\subsection{Proof of claims in Section E}
\subsubsection{Proof of Proposition E1}
\begin{flushleft}
\textbf{Proof of Proposition 3}: The proof of Proposition E1 follows that of Proposition 1 almost exactly and so only a sketch is provided here. Let $d_{ij;z}$ shorthand $||f_{w_{i};z}-f_{w_{j};z}||_{2}$ and $u_{i} = y_{i} - X_{i}\theta = \lambda(w_{i}) + \varepsilon_{i}$. I first demonstrate claim (ii). That $\lambda(w_{i}) = E\left[\left(y_{i} - X_{i}\theta\right) |\hspace{1mm}f_{w_{i};z}\right]$ follows from $E\left[\varepsilon|x_{i},w_{i}\right] = 0$ and Assumption E3 since 
\begin{align*}
E\left[u_{i}|f_{w_{i};z}\right] = E\left[\lambda(w_{i})|f_{w_{i};z}\right] + E\left[E\left[\varepsilon_{i}|X_{i}, w_{i}\right]|f_{w_{i};z}\right] = \lambda(w_{i}).
\end{align*}

I now demonstrate claim (i). That $\theta$ is the unique minimizer of $E\left[\left(y_{i}-y_{j} - (X_{i}-X_{j})b\right)^{2}|\hspace{1mm}||f_{w_i;z}-f_{w_j;z}||_{2} = 0\right]$ over $b \in \mathbb{R}^{2k+1}$ follows from expanding the square:
\begin{align*}
&E\left[\left(y_{i}-y_{j} - (X_{i}-X_{j})b\right)^{2}|d_{ij;z} = 0\right] = E\left[\left((X_{i}-X_{j})(\theta-b) + (u_{i}-u_{j})\right)^{2}|d_{ij;z} = 0\right]\\
&= (\theta-b)'E[(X_{i}-X_{j})'(X_{i}-X_{j}) |d_{ij;z} = 0](\theta - b) + E[(u_{i}-u_{j})^{2}|d_{ij;z} = 0] \\
&\hspace{20mm} -2(\theta - b)'E[(X_{i}-X_{j})'(u_{i}-u_{j}) |d_{ij;z} = 0].
\end{align*}
The first summand is uniquely minimized at $b = \theta$ by Assumption E2, the second summand does not depend on $b$, and the third summand is equal to $0$ by Assumption E3 and the assumption that $E\left[\varepsilon|x_{i},w_{i}\right] = 0$. $\square$
\newline
\end{flushleft}

\subsubsection{Proof of Proposition E2}
The proof of proposition E2 requires analogues of Lemma 1 and B1 for conditional link functions. These are given by Lemmas E1 and E2 respectively. 

\begin{flushleft} 
\textbf{Lemma E1}: Suppose Assumption D1. Then for any $i, j \in \{1,...,n\}$
\begin{align*}
||p_{w_{i};z} - p_{w_{j};z}||_{2} \leq ||g||_{2}^{2}||f_{w_{i};z} - f_{w_{j};z}||_{2}
\end{align*}
where $ ||g||_{2}^{2} = \int\int\int g(\tau,s;\zeta)^{2}d\tau ds d\zeta$ and for any $\varepsilon >0$ there exists a $\delta > 0$ such that 
\begin{align*}
||f_{w_{i;z}} - f_{w_{j;z}}||_{2}\times \mathbbm{1}\{||p_{w_{i;z}} - p_{w_{j;z}}||_{2} \leq \delta\} \leq \varepsilon.
\end{align*}
\end{flushleft} 

\begin{flushleft}
\textbf{Proof of Lemma E1}:
The proof of Lemma A3 mirrors that of Lemma 1. The first claim that $||p_{w_{i};z} - p_{w_{j};z}||_{2} \leq ||f_{w_{i};z} - f_{w_{j};z}||_{2}$ for any $i, j \in \{1,...,n\}$ follows from 
\begin{align*}
||p_{w_{i};z} - p_{w_{j};z}||_{2}^{2} &= \int\left(\int\int\left(f_{z}(\tau,s;\zeta)\left(f_{z}(w_{i},s;\zeta)- f_{z}(w_{j},s;\zeta)\right)\right)g(\tau,s;\zeta)dsd\zeta\right)^{2}d\tau  \\
&\leq \int\int\int\left(\left(f_{z}(\tau,s;\zeta)\left(f_{z}(w_{i},s;\zeta)- f_{z}(w_{j},s;\zeta)\right)\right)\right)^{2}dsd\zeta d\tau ||g||_{2}^{2}  \\
&\leq \int\int \left(f_{z}(w_{i},s;\zeta)-f_{z}(w_{j},s;\zeta)\right)^{2}dsd\zeta =||f_{w_{i};z} - f_{w_{j};z}||_{2}^{2}
\end{align*}
\end{flushleft}
where the first inequality is due to Cauchy-Schwarz and the fact that $g$ is square-integrable by Assumption D1 and the second inequality is due to the fact that $||f_{z}||_{\infty} \leq 1$. \newline

I now show that if $f_{z}$ satisfies Assumption D1 then for any $i, j \in \{1,...,n\}$ and $\varepsilon > 0$ there exists a $\delta > 0$ such that $||f_{w_{i};z}-f_{w_{j};z}||_{2}\mathbbm{1}\{||p_{w_{i};z}-p_{w_{j};z}||_{2} < \delta\} \leq \varepsilon$. Specifically I demonstrate the contrapositive: for any $i, j \in \{1,...,n\}$ and $\varepsilon > 0$ there exists a $\delta > 0$ such that $||p_{w_{i};z}-p_{w_{j};z}||_{2}\mathbbm{1}\{||f_{w_{i};z}-f_{w_{j};z}||_{2} > \varepsilon\} \geq \delta\mathbbm{1}\{||f_{w_{i};z}-f_{w_{j};z}||_{2} > \varepsilon\}$.  The two statements are equivalent because they are both violated if and only if $||p_{w_{i};z}-p_{w_{j};z}||_{2} < \delta$ and $||f_{w_{i};z}-f_{w_{j};z}||_{2} > \varepsilon$.

Fix $i,j \in \{1,...,n\}$, $\varepsilon >0$ and set $\underline{g} := \inf_{(u,v) \in [0,1]^{2}, \zeta \in \mathcal{Z}}g(u,v;\zeta)$ (which is positive by Assumption D1). Then 
\begin{align*}
||f_{w_i;z}-&f_{w_j;z}||_{2}^{2} = \int\int\left(f_{z}(w_i,\tau;\zeta)-f_{z}(w_j,\tau;\zeta)\right)^{2}d\tau d\zeta > \varepsilon \\
&\implies \int\int f_{z}(w_i,\tau;\zeta)(f_{z}(w_i,\tau;\zeta)-f(w_j,\tau;\zeta))d\tau d\zeta - \int\int f_{z}(w_j,\tau;\zeta)(f_{z}(u,\tau;\zeta)-f_{z}(w_j,\tau;\zeta))d\tau d\zeta  > \varepsilon \\
&\implies \left| \int\int f_{z}(x,\tau;\zeta)(f_{z}(w_i,\tau;\zeta)-f_{z}(w_j,\tau;\zeta))d\tau d\zeta \right| > \varepsilon/2 \text{ for some } x \in \{w_{i},w_{j}\} \\
&\implies \left| \int\int f_{z}(x,\tau;\zeta)(f_{z}(w_i,\tau;\zeta)-f_{z}(w_j,\tau;\zeta))g(x,\tau:\zeta)d\tau d\zeta \right| > \varepsilon \underline{g}/2 \text{ for some } x \in \{w_{i},w_{j}\} \\
&\implies \left|\int\int f_{z}(y,\tau;\zeta)(f_{z}(w_i,\tau;\zeta)-f_{z}(w_j,\tau;\zeta))g(x,\tau:\zeta)d\tau d\zeta\right| \\
&\hspace{20mm}+ \left|\int\int \left( f_{z}(x,\tau;\zeta)-f_{z}(y,\tau;\zeta)\right)(f_{z}(w_i,\tau;\zeta)-f_{z}(w_j,\tau;\zeta))g(x,\tau:\zeta)d\tau d\zeta\right| > \varepsilon\underline{g}/2 \\
&\hspace{20mm}\text{ for the } x \text{ above and any } y \in [0,1] \text{ by the triangle inequality}. \\
&\implies \left|\int\int f_{z}(y,\tau;\zeta)(f_{z}(w_i,\tau;\zeta)-f_{z}(w_j,\tau;\zeta))g(x,\tau:\zeta)d\tau d\zeta \right| > \varepsilon\underline{g}/4 \text{ for any } y \text{ such that } \\
&\hspace{20mm} \left|\int\int (f_{z}(x,\tau;\zeta)-f_{z}(y,\tau;\zeta)(f_{z}(w_i,\tau;\zeta)-f_{z}(w_j,\tau;\zeta))g(x,\tau:\zeta)d\tau \right| \leq \varepsilon\underline{g}/4 \text{ for the } x \text{ above.}
\end{align*}

By the continuity condition of Assumption D1, for every $u \in [0,1]$ and $\varepsilon' > 0$ the set $S(u,\varepsilon) := \{v \in [0,1], \zeta \in \mathcal{Z} : \left|f_{z}(u,\tau;\zeta)-f_{z}(v,\tau;\zeta)\right| \leq \varepsilon'\}$ has positive measure. Furthermore, since $[0,1]$ is compact, it must also be the case that $\omega(\varepsilon') := \inf_{u \in [0,1]}\left|S(u,\varepsilon')\right| > 0$ where $|\cdot|$ refers to the Lebesgue measure. Let $\overline{g} := \sup_{(u,v) \in [0,1]^{2}, \zeta \in \mathcal{Z}}g(u,v;\zeta)$ which is bounded since $g$ is a smooth function on a compact set by Assumption D1. It follows that choosing $y \in S(x,\varepsilon\underline{g}/4\overline{g})$ for the above $x$ implies that 
\begin{align*}
\left|\int\int (f_{z}(x,\tau;\zeta)-f_{z}(y,\tau;\zeta)(f_{z}(w_i,\tau;\zeta)-f_{z}(w_j,\tau;\zeta))g(x,\tau:\zeta)d\tau \right| \leq \varepsilon\underline{g}/4
\end{align*}
by the Cauchy-Schwartz inequality and the fact that $\left(f_{z}(w_{i},\tau)-f_{z}(w_{j},\tau)\right)g(z,\tau;\zeta)$ is absolutely bounded by $\overline{g}$. 

Consequently, 
\begin{align*}
||f_{w_i;z}-f_{w_j;z}||_{2}^{2} > \varepsilon &\implies \left|\int\int f_{z}(y,\tau;\zeta)(f_{z}(w_i,\tau;\zeta)-f_{z}(w_j,\tau;\zeta))g(y,\tau;\zeta)d\tau d\zeta \right| > \varepsilon/4 \\
&\hspace{20mm}\text{ for the } x  \text{ above and all } y \in S(x,\varepsilon\underline{g}/4\overline{g})  \\
&\implies ||p_{w_i;z}-p_{w_j;z}||_{2}^{2} = \int\left[\int\int f_{z}(y,\tau;\zeta)(f_{z}(w_i,\tau;\zeta)-f_{z}(w_j,\tau;\zeta))g(y,\tau;\zeta)d\tau d\zeta \right]^{2}dy \\
&\hspace{20mm}>  \frac{\varepsilon^{2}}{16}\times \omega(\varepsilon/4).
\end{align*}
It follows that for any $i,j \in \{1,...,n\}$ and $\varepsilon > 0$, $||p_{w_{i}}-p_{w_{j}}||_{2}\mathbbm{1}\{||f_{w_{i}}-f_{w_{j}}||_{2} > \varepsilon\} \geq \frac{\varepsilon^{2}}{16}\times \omega(\varepsilon\underline{g}/4\overline{g})$, and so $||f_{w_{i}}-f_{w_{j}}||_{2}\mathbbm{1}\{||p_{w_{i}}-p_{w_{j}}||_{2} < \delta\} \leq \varepsilon$ with $\delta = \frac{\varepsilon^{2}}{16}\times \omega(\varepsilon\underline{g}/4\overline{g})$. The claim follows. $\square$ \\

\begin{flushleft}
\textbf{Lemma E2}: Suppose Assumptions 4-5 and 8. Then
\begin{align*}
\max_{i \neq j}\left| \hat{\delta}^{2}_{ij;z} - ||p_{w_{i};z}- p_{w_{j};z}||^{2}_{2} \right| = o_{p}\left(n^{-\gamma/4}h_{n}\right)
\end{align*}
\end{flushleft}

\begin{flushleft}
\textbf{Proof of Lemma E2}:
The proof of Lemma A4 mirrors that of Lemma A1. Let $h_{n}' = n^{-\gamma/4}h_{n}$, $p_{w_tw_i;z} = \int\int f_{w_{t};z}(\tau;\zeta)f_{w_{i};z}(\tau;\zeta)g(w_{t},\tau;\zeta)d\tau d\zeta$, $\hat{p}_{w_{t}w_{i}w_{j};z} = \frac{\sum_{s\neq i,j,t}D_{ts}D_{is}K_{z}\left(\frac{|z_{is}-z_{ts}| + |z_{js} - z_{ts}|}{h_{n;z}}\right)}{\sum_{s\neq i,j,t}K_{z}\left(\frac{|z_{is}-z_{ts}| + |z_{js} - z_{ts}|}{h_{n;z}}\right)}$, $||\hat{p}_{w_{i};z} - p_{w_{i};z}||^{2}_{2,n,j} = (n-2)^{-1}\sum_{t \neq i,j}\left(\hat{p}_{w_{t}w_{i}w_{j};z} - p_{w_{t}w_{i};z}\right)^{2}$, and $||p_{w_{i};z}-p_{w_{j};z}||^{2}_{2,n} = (n-2)^{-1}\sum_{t \neq i,j}\left(p_{w_{t}w_{i};z} - p_{w_{t}w_{j};z}\right)^{2}$. Then for any fixed $\epsilon > 0$,
\begin{align*}
&P\left(\max_{i \neq j}h_{n}'^{-1} \left| \hat{\delta}_{ij;z}^{2} - ||p_{w_{i};z}-p_{w_{j};z}||^{2}_{2}  \right| > \epsilon \right) \\
&= P\left(\max_{i \neq j} h_{n}'^{-1} \left| \hat{\delta}_{ij;z}^{2} - ||p_{w_{i};z}-p_{w_{j};z}||^{2}_{2,n} + ||p_{w_{i};z}-p_{w_{j};z}||^{2}_{2,n} - ||p_{w_{i};z}-p_{w_{j};z}||^{2}_{2}  \right| > \epsilon \right) \\
&\leq P\left(\max_{i \neq j}h_{n}'^{-1} \left| \hat{\delta}^{2}_{ij;z} - ||p_{w_{i};z}-p_{w_{j};z}||^{2}_{2,n} \right|  > \epsilon/2 \right)  \\
&\hspace{30 mm} + P\left(\max_{i \neq j} h_{n}'^{-1}\left| ||p_{w_{i};z}-p_{w_{j};z}||^{2}_{2,n} - ||p_{w_{i};z}-p_{w_{j};z}||^{2}_{2}  \right|  > \epsilon/2 \right) \\
&= P\left(\max_{i \neq j}h_{n}'^{-1} \left| \hat{\delta}^{2}_{ij;z} - ||p_{w_{i};z}-p_{w_{j};z}||^{2}_{2,n} \right|  > \epsilon/2 \right) + o(1)  \\
&\leq P\left(\max_{i \neq j} h_{n}'^{-1}\left|(n-2)^{-1}\sum_{s \neq i,j}\left((\hat{p}_{w_{t}w_{i}w_{j}}-\hat{p}_{w_{t}w_{j}w_{i}}) -(p_{w_{t}w_{i}}-p_{w_{t}w_{j}})\right)\right|  > \epsilon/8 \right) + o(1) \\
&\leq 2P\left(\max_{i \neq j} h_{n}'^{-1}(n-2)^{-1}\sum_{t \neq i,j}\left|\hat{p}_{w_{t}w_{i}w_{j}} - p_{w_{t}w_{i}}\right|  > \epsilon/16 \right) + o(1) = o(1)
\end{align*}
\newline 
in which $P\left(\max_{i \neq j} h_{n}'^{-1}\left| ||p_{w_{i};z}-p_{w_{j};z}||^{2}_{2,n} - ||p_{w_{i};z}-p_{w_{j};z}||^{2}_{2}  \right|  > \epsilon/2 \right) = o(1)$ in the second equality and $P\left(\max_{i \neq j} h_{n}'^{-1}(n-2)^{-1}\sum_{t \neq i,j}\left|\hat{p}_{w_{t}w_{i}w_{j};z} - p_{w_{t}w_{i};z}\right|  > \epsilon/16 \right) = o(1)$ in the final equality are demonstrated below, the first inequality is due to the triangle inequality, the second inequality is due to the fact that $\left|p_{w_{i};z} + \hat{p}_{w_{i};z}\right| \leq 2$ for every $w_i \in [0,1]$, and the final inequality is due to the triangle and Jensen's inequality. 
\newline

The second result, that  $P\left(\max_{i \neq j} h_{n}'^{-1}(n-2)^{-1}\sum_{t \neq i,j}\left|\hat{p}_{w_{t}w_{i}w_{j};z} - p_{w_{t}w_{i};z}\right|  > \epsilon/16 \right) = o(1)$ follows from the fact that $\max_{i \neq j}h_{n}'^{-1}|\hat{p}_{w_{t}w_{i}w_{j};z}-p_{w_{t}w_{i};z}| \to_{p} 0$ by Bernstein's inequality and the union bound. Specifically, the former implies that for any $\epsilon > 0$
\begin{align*}
P&\left(\left|\left((n-3)h_{n;z}^{2L}\right)^{-1}\sum_{s\neq i,j,t}\left(\hat{A}_{ijt} - A_{ijt}\right)\right| > \epsilon/2\right)+P\left(\left|\left((n-3)h_{n;z}^{2L}\right)^{-1}\sum_{s\neq i,j,t}\left(\hat{B}_{ijt} - B_{ijt}\right)\right| > \epsilon/2\right) \\
&\leq 4\exp\left(\frac{-(n-3)h_{n;z}^{2L}\epsilon^{2}}{8 + 8\bar{K}_{z}\epsilon/3}\right)
\end{align*}
where $\bar{K_{z}} = \sup_{u}K_{z}(u)$ is finite by Assumption 8, $\hat{A}_{ijt} = K_{z}\left(\frac{|z_{is}-z_{ts}| + |z_{js}-z_{ts}|}{h_{n;z}}\right)$, $A_{ijt} = E\left[ K_{z}\left(\frac{|z_{is}-z_{ts}| + |z_{js}-z_{ts}|}{h_{n;z}} \right)\right]$, $\hat{B}_{ijt} = D_{ts}D_{is}K_{z}\left(\frac{|z_{is}-z_{ts}| + |z_{js}-z_{ts}|}{h_{n;z}}\right)$, and $B_{ijt} = E\left[D_{ts}D_{is}K_{z}\left(\frac{|z_{is}-z_{ts}| + |z_{js}-z_{ts}|}{h_{n;z}} \right)\right]$. Since $\min_{i,j,t}A_{ijt}$ is eventually bounded away from $0$ by Assumptions 5 and 8,
\begin{align*}
P&\left(\max_{i \neq j}|\hat{p}_{w_{t}w_{i}w_{j};z}-p_{w_{t}w_{i};z}| > \epsilon/2\right) \leq P\left(\max_{i\neq j \neq t}\left| \frac{\hat{B}_{ijt}}{\hat{A}_{ijt}} - \frac{B_{ijt}}{A_{ijt}} \right| > \varepsilon/2 \right) + o(1) \\
&\leq P\left(\left((n-3)h_{n;z}^{2L}\right)^{-1}\max_{i\neq j \neq t}\frac{1}{\hat{A}_{ijt}}\left| \hat{B}_{ijt} - B_{ijt} \right| > \varepsilon/2 \right)\\ 
&\hspace{5mm}+ P\left(\left((n-3)h_{n;z}^{2L}\right)^{-1}\max_{i\neq j \neq t}\frac{\hat{B}_{ijt}}{A_{ijt}\hat{A}_{ijt}}\left| \hat{A}_{ijt} - A_{ijt} \right| > \varepsilon/2 \right) + o(1) \\
& \leq 4n^{3}\exp\left(\frac{-(n-3)h_{n;z}^{2L}\epsilon^{2}}{8 + 8\bar{K}_{z}\epsilon^{3}/3}\right) + o(1)
\end{align*}
where $P\left(\max_{i\neq j \neq t}\left|\frac{B_{ijt}}{A_{ijt}}-p_{w_{t}w_{i};z}\right| > \epsilon/2 \right)  = o(1)$ in the first inequality follows from Assumption 5 and the last inequality is due to Bernstein's inequalitty the union bound. It follows that 
\begin{align*}
P\left(\max_{i \neq j} h_{n}'^{-1}(n-2)^{-1}\sum_{t \neq i,j}\left|\hat{p}_{w_{t}w_{i}w_{j};z} - p_{w_{t}w_{i};z}\right|  > \epsilon/16 \right) \leq 4n^{3}\exp\left(\frac{-(n-3)h_{n;z}^{2L}\left(h_{n}'\epsilon\right)^{2}}{2^{11}(1 + \bar{K}_{z}\left(h_{n}'\epsilon\right)/3)}\right)
\end{align*}
eventually, which is $o(1)$ since $nh_{n;z}^{2L+1}\to \infty$ for some $\gamma > 0$.
\newline 

The first result, that $P\left(\max_{i \neq j} h_{n}'^{-1}\left| ||p_{w_{i};z}-p_{w_{j};z}||^{2}_{2,n} - ||p_{w_{i};z}-p_{w_{j};z}||^{2}_{2}  \right|  > \epsilon/2 \right) = o(1)$, also follows from Bernstein's inequality and the union bound since
\begin{align*}
&P\left(h_{n}'^{-1}\left| ||p_{w_{i};z}-p_{w_{j};z}||^{2}_{2,n} - ||p_{w_{i};z}-p_{w_{j};z}||^{2}_{2} \right| > \epsilon \right) \\
& = P\left(h_{n}'^{-1}\left|(n-2)^{-1}\sum_{t \neq i,j}\left(p_{w_{t}w_{i};z} - p_{w_{t}w_{j};z}\right)^{2} - \int \left(p_{w_{i};z}(s)-p_{w_{j};z}(s)\right)^{2}ds\right| > \epsilon \right) \\
&\leq 2\exp\left(\frac{-(n-2)h_{n}'\epsilon}{2 + 2\sqrt{h_{n}'\epsilon}/3} \right) 
\end{align*}
which is $o(1)$ since $nh_{n}' \to \infty$. This completes the proof. $\square$ \newline
\end{flushleft}

\begin{flushleft}
\textbf{Proof of Proposition E2}: The proof of Proposition 4 follows that of Proposition 2 almost exactly, except Lemmas E1 and E2 are used instead of Lemmas 1 and B1. Consequently, only a sketch is provided here. \newline

First note that $\max_{i}\left|X_{i}-\hat{X}_{i}\right| = o_{p}\left(\log(n)/\sqrt{n}\right)$ by the usual uniform law of large numbers because it is a continuous function of $4n$ sample averages, each of independent mean zero entries. Write $u_{i} = y_{i} - \hat{X}_{i}\theta = (X_{i}-\hat{X}_{i})\theta + \lambda(w_{i}) + \varepsilon_{i} = o_{p}\left(\log(n)/\sqrt{n}\right) + \lambda(w_{i}) + \varepsilon_{i} $. I first consider $\hat{\theta}$, writing
\begin{align*}
\hat{\theta} = \theta + &\left( \sum_{i=1}^{n-1}\sum_{j =  i+1}^{n}(\hat{X}_{i}-\hat{X}_{j})'(\hat{X}_{i}-\hat{X}_{j})K\left(\frac{\hat{\delta}^{2}_{ij;z}}{h_{n}}\right)\right)^{-1}\left(\sum_{i=1}^{n-1}\sum_{j = i + 1}^{n}(\hat{X}_{i}-\hat{X}_{j})'(u_{i}-u_{j})K\left(\frac{\hat{\delta}^{2}_{ij;z}}{h_{n}}\right)\right).
\end{align*}
The denominator  $\left({n \choose 2}r_{n}\right)^{-1}\sum_{i=1}^{n-1}\sum_{j =  i+1}^{n}(\hat{X}_{i}-\hat{X}_{j})'(\hat{X}_{i}-\hat{X}_{j})K\left(\frac{\hat{\delta}^{2}_{ij;z}}{h_{n}}\right) \to_{p} r_{n}^{-1}E\left[\left(X_{i}-X_{j}\right)'(X_{i}-X_{j})K\left(\frac{\delta_{ij;z}^{2}}{h_{n}}\right)\right]$ following exactly the arguments of Proposition 2 and Lemmas E1 and E2. The numerator $\left({n \choose 2}r_{n}\right)^{-1}\sum_{i=1}^{n-1}\sum_{j =  i+1}^{n}(\hat{X}_{i}-\hat{X}_{j})'(u_{i}-u_{j})K\left(\frac{\hat{\delta}^{2}_{ij;z}}{h_{n}}\right)$ is similarly equal to
\begin{align*}
\left({n \choose 2}r_{n}\right)^{-1}\sum_{i=1}^{n-1}\sum_{j =  i+1}^{n}(X_{i}-X_{j})'\left[(\varepsilon_{i}-\varepsilon_{j}) + (\lambda(w_{i})-\lambda(w_{j}))\right]K\left(\frac{\hat{\delta}^{2}_{ij;z}}{h_{n}}\right) + o_{p}\left(\sqrt{\log(n)/n}\right)
\end{align*}
which is $o_{p}(1)$ following exactly the arguments of Proposition 2 and Assumption E3. The first summand is $o_{p}(1)$ because $X_{i}$ has bounded second moments and $E\left[\varepsilon_{i}|x_{i},w_{i}\right] = 0$. The result $\hat{\theta} \to_{p} \theta$ then follows since the denominator is uniformly bounded over $\mathbb{N}$ by Assumption E2, following exactly the logic of Proposition 2. \newline

Similarly it follows from the arguments in the second part of the proof of Proposition 2 and Lemma E2 that
\begin{align*}
\max_{i=1,...,n}\left| \frac{1}{n}\sum_{t=1}^{n}K\left(\frac{\hat{\delta}^{2}_{it;z}}{h_{n}}\right) - E\left[K\left(\frac{\delta^{2}_{it;z}}{h_{n}}\right)|w_{i}\right]\right| = o_{p}\left(n^{-\gamma/4}h_{n}\right),
\end{align*} 
\begin{align*}
\max_{i=1,...,n}\left| \frac{1}{n}\sum_{t=1}^{n}\left(y_{t} - \hat{X}_{t}\hat{\theta}\right)K\left(\frac{\hat{\delta}^{2}_{it;z}}{h_{n}}\right) - E\left[\left(y_{t}-\hat{X}_{t}\hat{\theta}\right)K\left(\frac{\delta^{2}_{it;z}}{h_{n}}\right)|w_{i}\right]\right| = o_{p}\left(n^{-\gamma/4}h_{n}\right)
\end{align*}
and so by the continuous mapping theorem and the fact that $\min_{i=1,...,n}E\left[K\left(\frac{\delta^{2}_{it;z}}{h_{n}}\right)|w_{i}\right]$ is eventually bounded away from zero by the choice of kernel in Assumption 4
\begin{align*}
\max_{i=1,...,n}\left|\widehat{\lambda(w_{i})} - \frac{E\left[\lambda(w_{t})K\left(\frac{\delta_{it;z}}{h_{n}}\right)|w_{i}\right]}{E\left[K\left(\frac{\delta_{it;z}}{h_{n}}\right)|w_{i}\right]} +   \frac{E\left[\hat{X}_{t}K\left(\frac{\delta_{it;z}}{h_{n}}\right)|w_{i}\right]\left(\hat{\theta} - \theta\right)}{E\left[K\left(\frac{\delta_{i;zt}}{h_{n}}\right)|w_{i}\right]}\right| = o_{p}\left(n^{-\gamma/4}h_{n}\right)
\end{align*}

Since $X_{i}$ has finite second moments, $\frac{E\left[\hat{X}_{t}K\left(\frac{\delta_{it;z}}{h_{n}}\right)|w_{i}\right]}{E\left[K\left(\frac{\delta_{it;z}}{h_{n}}\right)|w_{i}\right]}$ is uniformly bounded and so $\max_{i = 1,...,n} \frac{E\left[\hat{X}_{t}K\left(\frac{\delta_{it;z}}{h_{n}}\right)|w_{i}\right]\left(\hat{\theta} - \theta\right)}{E\left[K\left(\frac{\delta_{it;z}}{h_{n}}\right)|w_{i}\right]} = o_{p}(1)$ by previous arguments. The claim then follows since $\max_{i = 1,...,n}\left|\lambda(w_{i})-  \frac{E\left[\lambda(w_{t})K\left(\frac{\delta_{it};z}{h_{n}}\right)|w_{i}\right]}{E\left[K\left(\frac{\delta_{it};z}{h_{n}}\right)|w_{i}\right]} \right| = o_{p}(1)$ by Assumption E4 and Lemma E2, following exactly the logic of Proposition 2. $\square$
\end{flushleft}

\section{Simulation evidence}
This appendix presents simulation evidence for three types of network formation models described in Section 2.2.1: a stochastic blockmodel, a degree heterogeneity model, and a homophily model. For each of $R$ simulations, I draw a random sample of $n$ observations $\{\xi_{i},\varepsilon_{i},\omega_{i}\}_{i=1}^{n}$ from a trivariate normal distribution with mean $0$ and covariance given by the identity matrix and a random symmetric matrix $\{\eta_{ij}\}_{i,j =1 }^{n}$ with independent and identically distributed upper diagonal entries with standard uniform marginals. For each of the following link functions $f$, the adjacency matrix $D$ is formed by $D = \mathbbm{1}\{\eta_{ij} \leq f\left(\Phi(\omega_i),\Phi(\omega_j)\right)\}$ where $\Phi$ is the cumulative distribution function for the standard univariate normal distribution. 

The first design draws $D$ from a stochastic blockmodel where
\[f_{1}(u,v) =   \left\{ \begin{array}{cccc}
1/3 &&\text{ if } u \leq 1/3 \text{ and } v > 1/3 \\
1/3 &&\text{ if } 1/3 < u \leq 2/3 \text{ and } v \leq 2/3 \\
1/3 &&\text{ if } u > 2/3 \text{ and } (v > 2/3 \text{ or } v \leq 1/3) \\
0 &&\text{ otherwise } \end{array}
 \right.
\]
The linking function $f_{1}$ generates network types with finite support as in the hypothesis of Proposition C1. For this model, I take $\lambda(\omega_{i}) = \lceil 3\Phi(\omega_{i})\rceil$, $x_{i} = \xi_{i} + \lambda(\omega_{i})$, and $y_{i} = \beta x_{i} + \gamma \lambda(\omega_{i}) + \varepsilon_{i}$. The second and third designs draw $D$ from the degree heterogeneity model and homophily model where 
\[f_{2}(u,v) = \frac{\exp(u+v)}{1+\exp(u+v)} \text{ and }f_{3}(u,v) = 1-(u-v)^2\]
For these models, $\lambda(\omega_{i}) = \omega_{i}$, $x_{i} = \xi_{i} + \lambda(\omega_{i})$ and $y_{i} = \beta x_{i} + \gamma \lambda(\omega_{i}) + \varepsilon_{i}$. 

Let $x$ and $y$ to denote the stacked $n$-dimensional vector of observations $\{x_{i}\}_{i=1}^{n}$ and $\{y_{i}\}_{i=1}^{n}$, and $Z_{1}$ for the $(n \times 2)$ matrix $\{x_{i},\lambda(\omega_{i})\}_{i=1}^{n}$. I use $c_{i}$ to denote a vector of network statistics for agent $i$ based on $D$ containing agent degree $n^{-1}\sum_{j=1}^{n}D_{ij}$, eigenvector centrality,\footnote{Agent $i$'s eigenvector centrality statistics refers to the $i$th entry of the eigenvector of $D$ associated with the largest eigenvalue.} and average peer covariates $\sum_{j=1}^{n}D_{ij}x_{j}/\sum_{j=1}^{n}D_{ij}$. $Z_{2}$ denotes the stacked vector $\{x_{i}, c_{i}\}_{i=1}^{n}$.

For each design, I evaluate the performance of six estimators. The benchmark is $\hat{\beta}_{1} = (Z_{1}'Z_{1})^{-1}(Z_{1}'y)$, the infeasible OLS regression of $y$ on $x$ and $\lambda(\omega_{i})$. $\hat{\beta}_{2} = (x'x)^{-1}(x'y)$ is the na\"ive OLS regression of $y$ on $x$. $\hat{\beta}_{3} = (Z_{2}'Z_{2})^{-1}(Z_{2}'y)$ is the OLS regression of $y$ on $x$ and the vector of network controls $c$. $\hat{\beta}_{4}$ is the proposed pairwise difference estimator given in (4) without bias correction, $\hat{\beta}_{5}$ is the bias corrected estimator, and $\hat{\beta}_{6}$ is the pairwise difference estimator with an adaptive bandwidth but without bias correction (specifically, the bandwidth depends on $i$ and is chosen such that each agent is matched to the same number of other agents). The pairwise difference estimators all use the Epanechnikov kernel $K(u) = 3(1-u^{2})\mathbbm{1}\{u^{2} < 1\}/4$. Estimators $\hat{\beta}_{4}$ and $\hat{\beta}_{5}$ use the  bandwidth sequence $n^{-1/9}/10$ and the estimator $\hat{\beta}_{6}$ uses the bandwidth sequence $n^{-1/9}/5$. Since $n^{1/9}$ is roughly equal to $2$ for the sample sizes considered in this section, the results are close to a constant bandwidth choice of $h_{n} = .05$ and $.1$ respectively.  

Tables 1-3 demonstrates the results for $R = 1000$, $\beta = \gamma = 1$ and for each $n$ in $\{50,100,200,500, 800\}$. For each model, estimator and sample size, the first row gives the mean, the second gives the mean absolute error of the simulated estimators around $\beta$, the third gives the mean absolute error divided by that of $\hat{\beta}_{1}$, and the fourth gives the proportion of the simulation draws that fall outside of a $0.95$ confidence interval based on the asymptotic distributions derived in the previous section.

\begin{table}[]
\centering
\title{Table 1: Simulation Results, Stochastic Blockmodel} \\ \vspace{5mm}
\label{Table 1}
\begin{tabular*}{\textwidth}{c @{\extracolsep{\fill}} cccccccc}
\toprule \toprule 
&& Infeasible  & Na\"ive  & OLS with  & Pairwise  & Bias & Adaptive \\ 
&& OLS  & OLS  & Controls  & Difference  & Corrected & Bandwidth \\ 
n&&$\hat{\beta}_{1}$ & $\hat{\beta}_{2}$ & $\hat{\beta}_{3}$ & $\hat{\beta}_{4}$ & $\hat{\beta}_{5}$ &$\hat{\beta}_{6}$ \\ \midrule
50&&      &     & & & &      \\
&bias& 0.004  & 0.829        & 0.268    & 0.060& 0.022& 0.106   \\
&MAE&  0.116   & 0.829     & 0.274    & 0.224& 0.240 & 0.150  \\
&rMAE&  1.000  & 7.147 & 2.362 & 1.931 & 2.069& 1.293 \\
&size&  0.057   & 0.063     & 0.072    & 0.115 & 0.123& 0.067  \\ 
  
100&&      &     & & & &      \\
&bias& 0.003  & 0.829        & 0.226    & 0.021& -0.022& 0.019   \\
 &MAE&  0.083   & 0.829     & 0.229    & 0.089& 0.094 & 0.084  \\
 &rMAE& 1.000 &  9.988 & 2.759 & 1.072 & 1.133 & 1.012 \\
  &size&  0.064   & 0.053     & 0.108    & 0.053 & 0.058& 0.056  \\ 
  200&&      &     & & & &      \\
&bias& 0.001  & 0.823        & 0.180   & 0.004& -0.040& 0.002   \\
 &MAE&  0.056   & 0.823     & 0.183    & 0.058& 0.069 & 0.058  \\
  &rMAE& 1.000 & 14.696 & 3.268 & 1.036 & 1.232 & 1.036 \\
  &size&  0.049   & 0.044     & 0.215    & 0.045 & 0.064& 0.058  \\ 
    500&&      &     & & & &      \\
&bias& 0.000  & 0.824       & 0.172   & 0.006& 0.038& 0.001   \\
 &MAE&  0.035   & 0.824     & 0.174    & 0.035& 0.048 & 0.035  \\
  &rMAE& 1.000 & 23.543 & 4.971 & 1.000 & 1.371 & 1.000\\
  &size&  0.033   & 0.061     & 0.777    & 0.037 & 0.047& 0.044  \\ 
  
 800&&      &     & & & &      \\
&bias& 0.001  & 0.823        & 0.314    & 0.008& -0.036& 0.000   \\
 &MAE&  0.029   & 0.823     & 0.314    & 0.029& 0.043 & 0.029  \\
   &rMAE& 1.000 & 28.379 & 10.828 & 1.000 & 1.483 & 1.000\\
  &size&  0.057   & 0.038     & 0.127    & 0.054 & 0.068& 0.062 \\
  \bottomrule
\end{tabular*}
\caption{\footnotesize This table contains simulation results for $1000$ replications and a sample size of $n = 50,100, 200, 500,800$. Bias gives the mean estimator minus $1$. MAE gives the mean absolute error of the estimator around $1$. rMAE gives the mean absolute error relative to the benchmark $\hat{\beta}_{1}$. Size gives the proportion of draws that fall outside the asymptotic $0.95$ confidence interval.\normalsize}
\end{table}

Table 1 contains results for the stochastic blockmodel. The na\"ive estimator $\hat{\beta}_{2}$ has a large and stable positive bias that is not reduced as $n$ is increased. The OLS estimator with network controls $\hat{\beta}_{3}$ is not asymptotically well defined in this example because the network statistics converge to constants. The results in Table 1 instead demonstrate a common ``fix'' in the literature, which is to instead calculate  $(Z_{2}'Z_{2})^{+}(Z_{2}'y)$ where $+$ refers to the Moore-Penrose pseudo-inverse. The results for this estimator indicate that adding network controls mitigates some of the bias in $\beta_{1}$ (due to sampling variation in the number of agents in each block), however the estimator is otherwise poorly behaved. Notice this bias returns when the block sizes stabilize (in particular when $n = 800$). 

The results for the pairwise difference estimators illustrate the content of Proposition C1, that when the unobserved heterogeneity is discrete, the proposed estimator identifies pairs of agents of the same type with high probability. As a result, the pairwise difference estimators $\hat{\beta}_{4}$ and $\hat{\beta}_{6}$ behave similarly to the infeasible $\hat{\beta}_{2}$. For the stochastic blockmodel, Assumption C3 is not valid, and so the jackknife bias correction actually inflates both the bias and variance of $\hat{\beta}_{4}$. Looking at the relative mean absolute error for this estimator, it is clear that the relative performance of the error is deteriorating as $n$ increases (though the bias and variance of this estimator still appear to bel on the order of $1/\sqrt{n}$).    

\begin{table}[]
\centering
\title{Table 2: Simulation Results, Beta Model} \\ \vspace{5mm}
\label{Table 2}
\begin{tabular*}{\textwidth}{c @{\extracolsep{\fill}} cccccccc}
\toprule \toprule
&& Infeasible  & Na\"ive  & OLS with  & Pairwise  & Bias & Adaptive \\ 
&& OLS  & OLS  & Controls  & Difference  & Corrected & Bandwidth \\ 
n&&$\hat{\beta}_{1}$ & $\hat{\beta}_{2}$ & $\hat{\beta}_{3}$ & $\hat{\beta}_{4}$ & $\hat{\beta}_{5}$ &$\hat{\beta}_{6}$ \\ \midrule
50&&      &     & & & &      \\
&bias& 0.000  & 0.496        & 0.462    & 0.379& 0.335& 0.365   \\
&MAE&  0.119   & 0.496     & 0.463    & 0.381& 0.341 & 0.366  \\
&rMAE & 1.000 & 4.168 & 3.891 & 3.202 & 2.866 & 3.076 \\
&size&  0.064   & 0.063     & 0.075    & 0.049 & 0.066& 0.070  \\ 

100&&      &     & & & &      \\
&bias& 0.006  & 0.501        & 0.462    & 0.336& 0.269& 0.298   \\
 &MAE&  0.082   & 0.501     & 0.462    & 0.336& 0.270 & 0.299  \\
 &rMAE  & 1.000 & 6.110 &  5.634 & 4.098 & 3.293 & 3.646 \\
  &size&  0.055   & 0.053     & 0.055    & 0.039 & 0.062& 0.081  \\ 
  200&&      &     & & & &      \\
&bias& 0.002  & 0.501        & 0.444   & 0.290& 0.200& 0.231   \\
 &MAE&  0.058   & 0.501     & 0.444    & 0.290& 0.200 & 0.231  \\
 &rMAE  & 1.000 & 8.638 & 7.655 & 5.000& 3.448 & 3.983 \\
  &size&  0.050   & 0.041     & 0.036    & 0.033 & 0.054& 0.070  \\ 
    500&&      &     & & & &      \\
&bias& 0.003  & 0.499       & 0.403   & 0.246& 0.136 & 0.151   \\
 &MAE&  0.036   & 0.499     & 0.403    & 0.246& 0.136 & 0.151  \\
 &rMAE  & 1.000 & 13.861 & 11.194 & 6.833 & 3.778 & 4.194 \\ 
  &size&  0.049   & 0.042     & 0.054    & 0.022 & 0.033& 0.076  \\
 
 800&&      &     & & & &      \\
&bias& 0.000  & 0.500        & 0.385    & 0.237& 0.122& 0.122   \\
 &MAE&  0.028   & 0.500     & 0.385    & 0.237& 0.122 & 0.122  \\
 &rMAE  & 1.000 & 17.857 & 13.750& 8.464 & 4.357 & 4.357\\
  &size&  0.050   & 0.054     & 0.078    & 0.037 & 0.050& 0.062  \\  
   \bottomrule
\end{tabular*}
\caption{\footnotesize This table contains simulation results for $1000$ replications and a sample size of $n = 50,100, 200, 500,800$. Bias gives the mean estimator minus $1$. MAE gives the mean absolute error of the estimator around $1$. rMAE gives the mean absolute error relative to the benchmark $\hat{\beta}_{1}$. Size gives the proportion of draws that fall outside the asymptotic $0.95$ confidence interval.\normalsize}
\end{table}

Table 2 contains results for the degree heterogeneity model. Relative to the stochastic blockmodel, all of the estimators for the beta model (except infeasible OLS) have large biases. This is because the link function $f_{2}$ is very flat, so that the variation in linking probabilities that identifies the network positions is relatively small \cite*[see also Section 5 of][]{johnsson2015estimation}. As argued in Section 2.2.1, the social characteristics are identified by the distribution of $D$ (they are consistently estimated by the order statistics of the degree distribution), but the bound on the deviation of the social characteristics given by the network metric is large: $|u-v| \leq 20\times d(u,v)$. 

Still, the proposed pairwise difference estimator offers a substantial improvement in performance relative to both the na\"ive estimator $\hat{\beta}_{2}$ and the estimator with network controls $\hat{\beta}_{3}$. For example, when $n = 100$, $\hat{\beta}_{5}$ has approximately half the bias and mean absolute error of $\hat{\beta}_{2}$ while $\hat{\beta}_{3}$ offers a reduction of less than ten percent. When $n = 800$ the reduction in bias is over three times as large (75\% relative to 23\%).  

\begin{table}[]
\centering
\title{Table 3: Simulation Results, Homophily Model} \\ \vspace{5mm}
\label{Table 3}
\begin{tabular*}{\textwidth}{c @{\extracolsep{\fill}} cccccccc}
\toprule \toprule
&& Infeasible  & Na\"ive  & OLS with  & Pairwise  & Bias & Adaptive \\ 
&& OLS  & OLS  & Controls  & Difference  & Corrected & Bandwidth \\ 
n&&$\hat{\beta}_{1}$ & $\hat{\beta}_{2}$ & $\hat{\beta}_{3}$ & $\hat{\beta}_{4}$ & $\hat{\beta}_{5}$ &$\hat{\beta}_{6}$ \\ \midrule
50&&      &     & & & &      \\
&bias& 0.007  & 0.505        & 0.269    & 0.128& 0.087& 0.140   \\
 &MAE&  0.120   & 0.505     & 0.274    & 0.108& 0.121 & 0.211  \\
 &rMAE& 1.000 & 4.208 & 2.283 & 0.900 & 1.008 & 1.758 \\
  &size&  0.068   & 0.051     & 0.063    & 0.062 & 0.068& 0.132  \\ 

100&&      &     & & & &      \\
&bias& 0.005  & 0.502        & 0.162    & 0.100& 0.057& 0.089   \\
 &MAE&  0.081   & 0.502     & 0.167    & 0.124& 0.108 & 0.116  \\
 &rMAE& 1.000 & 6.198 & 2.062 & 1.531 & 1.333 & 1.432 \\
  &size&  0.049   & 0.059     & 0.061    & 0.053 & 0.066& 0.083  \\ 
  200&&      &     & & & &      \\
&bias& 0.001  & 0.503        & 0.095   & 0.085& 0.039& 0.055   \\
 &MAE&  0.057   & 0.503     & 0.100    & 0.097& 0.075 & 0.077  \\
 &rMAE& 1.000 & 8.825 & 1.754 & 1.702 & 1.316 & 1.351 \\
  &size&  0.054   & 0.059     & 0.054    & 0.050 & 0.057& 0.069  \\ 
    500&&      &     & & & &      \\
&bias& 0.000  & 0.501       & 0.047   & 0.074& 0.028 & 0.035   \\
 &MAE&  0.035   & 0.501     & 0.053    & 0.077& 0.048 & 0.046  \\
 &rMAE& 1.000 & 14.314 & 1.514 & 2.200 & 1.371 & 1.314 \\
  &size&  0.043   & 0.059     & 0.039    & 0.045 & 0.058& 0.051  \\
 
 800&&      &     & & & &      \\
&bias& 0.000  & 0.501        & 0.034    & 0.070& 0.023& 0.030   \\
 &MAE&  0.028   & 0.501     & 0.086    & 0.072& 0.039 & 0.038  \\
 &rMAE& 1.000 & 17.893 & 3.071 & 2.571 & 1.392 & 1.357 \\
  &size&  0.039   & 0.040     & 0.041    & 0.038 & 0.050& 0.047  \\  
   \bottomrule
\end{tabular*}
\caption{\footnotesize This table contains simulation results for $1000$ replications and a sample size of $n = 50,100, 200, 500,800$. Bias gives the mean estimator minus $1$. MAE gives the mean absolute error of the estimator around $1$. rMAE gives the mean absolute error relative to the benchmark $\hat{\beta}_{1}$. Size gives the proportion of draws that fall outside the asymptotic $0.95$ confidence interval.\normalsize}
\end{table}

Table 3 contains results for the homophily model. As in the case of the degree heterogeneity model, one can show that the network pseudometric is in fact a metric on $[0,1]$ (although the actual position of $w_{i}$ in $[0,1]$ is not knowable). Unlike the degree heterogeneity model, there is a relatively large amount of information about the network positions in the linking probabilities so that all of the estimators in Table 3 are much better behaved. In fact, for this model $|u-v| \leq d(u,v)$.  

In this example, the OLS estimator with network controls actually performs comparably to the uncorrected pairwise difference estimator $\hat{\beta}_{4}$. This is because the peer characteristics variable $\sum_{j=1}^{n}D_{ij}x_{j}/\sum_{j=1}^{n}D_{ij}$ is a good approximation of $w_{i}$ when $n$ is large. However, the bias corrected estimator $\hat{\beta}_{5}$ outperforms both estimators over all of the sample sizes considered.

\bibliographystyle{chicago}
\bibliography{literature}
\end{document}